\documentclass[12pt]{article}
\pdfoutput=1 

\usepackage[oldstyle]{hep-paper}
\usepackage[compat=1.1.0]{tikz-feynman}
\tikzset{graviton/.style={decorate, decoration={snake, amplitude=.4mm, segment length=1.5mm, pre length=.5mm, post length=.5mm}, double}}
\tikzfeynmanset{doublefermion/.style={
/tikz/double,
/tikz/decoration={name=none},
/tikz/postaction={
/tikzfeynman/with arrow=0.5,
}
}
}

\tikzfeynmanset{doubleantifermion/.style={
/tikz/double,
/tikz/decoration={name=none},
/tikz/postaction={
/tikzfeynman/with reversed arrow=0.5,
}
}
}

\usepackage{scalerel}							
\usepackage{subdepth}							
\usepackage{ulem}
\usepackage{subfiles}
\usepackage{import}
\usepackage{appendix}							
\AtBeginEnvironment{subappendices}{%
  \addtocontents{toc}{\protect\addvspace{10pt}Appendices}
}

\usepackage[british]{babel}						
\usepackage{upgreek}
\usepackage{xcolor}
\usepackage{verbatim}

\usepackage{tabularx}							%
\usepackage{makecell}							%
\usepackage[export = true]{adjustbox}				%

\usepackage{dsfont}

\usepackage{tikz}
\usepackage{tikzscale}
\usepackage{tikz-feynman}

\usepackage{bibentry}
\usepackage[nottoc]{tocbibind}

\allowdisplaybreaks[1] 							

\usepackage{tocloft}

\AtBeginEnvironment{tabular}{\tlstyle}

\newcommand{\makediff}[2]{\expandafter\NewDocumentCommand\csname#1\endcsname{ogd()}{\IfNoValueTF{##2}{\IfNoValueTF{##3}{#2\IfNoValueTF{##1}{}{^{##1}}}{\mathinner{#2\IfNoValueTF{##1}{}{^{##1}}\argopen(##3\argclose)}}}{\mathinner{#2\IfNoValueTF{##1}{}{^{##1}}##2} \IfNoValueTF{##3}{}{(##3)}}}}
\makediff{mD}{\mathcal D}
\DeclareMathOperator{\T}{T}

\DeclareMathOperator{\SU}{SU}
\mathdef{\U}{\operatorname{U}}
\mathdef{\P}{\operatorname{P}}

\newcommand{\sparallel}{ {\scaleto{\parallel}{5pt}} }

\acronym{QFT}{quantum field theory}[quantum field theories]
\acronym{QCD}{quantum chromodynamics}
\acronym{QED}{quantum electrodynamics}
\acronym{GD}{gluon dynamics}
\acronym{SM}{Standard Model}
\acronym{RGE}{renormalization group equation}
\acronym{VEV}{vacuum expectation value}
\acronym{CKM}{Cabibbo-\allowbreak Kobayashi-\allowbreak Maskawa}
\acronym{PMNS}{Pontecorvo–\allowbreak Maki–\allowbreak Nakagawa–\allowbreak Sakata}
\acronym[$\chi$PT]{cPT}{chiral perturbation theory}[chiral perturbation theories]
\acronym{NGB}{Nambu-\allowbreak Goldstone boson}
\acronym{PNGB}{pseudo Nambu-\allowbreak Goldstone boson}
\acronym{WZW}{Wess-\allowbreak Zumino-\allowbreak Witten}
\acronym{MC}{Maurer-\allowbreak Cartan}
\acronym{GM}{Gell-\allowbreak Mann}
\acronym{CS}{Chern-\allowbreak Simons}
\acronym{YM}{Yang-\allowbreak Mills}
\acronym{HNL}{heavy neutral lepton}
\acronym{ALP}{axion-like particle}
\acronym{LLP}{long-lived particle}
\acronym{EOM}{equation of motion}
\acronym{PI}{partial integration}
\acronym[1PI]{onePI}{one particle irreducible}
\acronym{BSM}{beyond the Standard Model}
\longacronym{WIMP}{weakly interacting massive particle}
\acronym{DM}{dark matter}
\acronym{UV}{ultraviolet}
\acronym{IR}{infrared}
\acronym{LEC}{low energy coefficient}
\acronym{LER}{low energy realisation}
\acronym{NP}{new physics}
\acronym{EM}{electromagnetic}
\acronym{EW}{electroweak}
\acronym{EWSB}{electroweak symmetry breaking}
\acronym[\overline{MS}]{MSb}{modified minimal subtraction}
\acronym{NDA}{naive dimensional analysis}
\acronym{DOF}{degree of freedom}[degrees of freedom]
\acronym{LO}{leading order}
\acronym{NLO}{next-to-leading order}
\acronym{NNLO}{next-to-next-to-leading order}
\acronym{LHC}{Large Hadron Collider}
\shortacronym{CMS}{compact muon solenoid}
\shortacronym{ATLAS}{a toroidal LHC apparatus}
\shortacronym{LHCb}{LHC beauty}
\shortacronym{NA}*{North Area experiment \textnumero~}
\shortacronym{KOTO}{K0 at Tokai}
\shortacronym{SHiP}{Search for Hidden Particles}
\shortacronym{MATHUSLA}{Massive Timing Hodoscope for Ultra-Stable neutraL pArticles}
\shortacronym{FASER}{ForwArd Search ExpeRiment}
\shortacronym[CODEX-b]{CODEXb}{compact detector for exotics at LHCb}
\shortacronym{CP}{charge-parity}
\acronym{PET}{\emph{portal effective theory}}[\emph{portal effective theories}]
\acronym{EFT}{effective field theory}[effective field theories]
\acronym{SMEFT}{Standard Model effective field theory}[Standard Model effective field theories]
\acronym{HEFT}{Higgs effective field theory}[Higgs effective field theories]
\acronym{LEFT}{light effective field theory}[light effective field theories]
\acronym{SCET}{soft-collinear effective theory}[soft-collinear effective theories]
\acronym{HQET}{heavy quark effective theory}[heavy quark effective theories]
\acronym{NRQCD}{non-relativistic quantum chromodynamics}
\acronym{NRQED}{non-relativistic quantum electrodynamics}
\acronym[2HDM]{THDM}{two Higgs-doublet model}
\acronym{SUSY}{supersymmetry}
\acronym{SUGRA}{supergravity}
\acronym{MSSM}{minimal supersymemtric Standard Model}
\acronym{NMSSM}{next-to-minimal supersymemtric Standard Model}
\longacronym{CC}{charged current}
\longacronym{NC}{neutral current}
\longacronym{LE}{low energy}[low energies]
\longacronym{HE}{high energy}[high energies]
\acronym{BR}{branching ratio}
\acronym{GW}{gravitational wave}
\acronym{GWB}{gravitational wave background}
\acronym{LISA}{Laser Interferometer Space Antenna}
\acronym{ET}{Einstein Telescope}
\acronym{DECIGO}{Deci-hertz Interferometer Gravitational wave Observatory}
\acronym{BBO}{Big Bang Observer}
\acronym{CMB}{cosmic microwave background}

\acronyms{Omit}

\acronym{SNF}{Schweizerischer Nationalfonds zur Förderung der wissenschaftlichen Forschung}
\acronym[exp]{ex}{experimental}
\acronym{lat}{lattice}
\acronym{UCLouvain}{Université catholique de Louvain}
\acronym{FSR}{Fonds Spéciaux de Recherche}

\crefname{type}{type}{types}
\crefalias{inlinelisti}{type}
\crefalias{enumi}{type}

\crefformat{footnote}{#2\footnotemark[#1]#3}

\eqcrefname{lag}{Lagrangian}
\eqcrefname{set}{stress-energy tensor}
\eqcrefname{cur}{current}
\eqcrefname{sym}{symmetry}
\eqcrefname{bilinear}{quark bilinear}

\usepackage{tocloft}

\AtBeginEnvironment{tabular}{\tlstyle}

\let\oldfigure\figure
\def\figure{\oldfigure\small}
\let\oldtable\table
\def\table{\oldtable\small}

\newcommand{\Uav}{\mathrm\Upsilon_\text{av}}

\makeatletter

\def\custom#1{{\hbox{$\left#1\vbox to30\p@{}\right.\n@space$}}}
\makeatother

\def\int{\intop\nolimits}

\graphicspath{{./figures/}}
\bibliography{bibliography}

\title{Upper Bound on Thermal Gravitational Wave Backgrounds from Hidden Sectors}

\author[louvain]{Marco Drewes\footnote{\texttt{marco.drewes@uclouvain.be}}}
\author[louvain]{Yannis Georis\footnote{\texttt{yannis.georis@uclouvain.be}}}
\author[uva, nikhef, zagreb, louvain]{Juraj Klaric\footnote{\texttt{juraj.klaric@nikhef.nl}}}
\author[bern,bielefeld]{Philipp Klose\footnote{\texttt{pklose@physik.uni-bielefeld.de}}} 

\affiliation[louvain]{Centre for Cosmology, Particle Physics, and Phenomenology, Université catholique de Louvain, Louvain-la-Neuve B-1348, Belgium}
\affiliation[uva]{Institute for Theoretical Physics Amsterdam and Delta Institute for Theoretical Physics,University of Amsterdam, Science Park 904, 1098 XH Amsterdam, The Netherlands}
\affiliation[nikhef]{Nikhef, Theory Group, Science Park 105, 1098 XG, Amsterdam, The Netherlands}
\affiliation[zagreb]{University of Zagreb, Faculty of Science, Department of Physics, 10000 Zagreb, Croatia}
\affiliation[bern]{Institute for Theoretical Physics, Albert Einstein Center for Fundamental Physics, University of Bern}
\affiliation[bielefeld]{Fakultät für Physik, Universität Bielefeld}

\begin{document}

\maketitle 

\begin{abstract}
Hot viscous plasmas unavoidably emit a gravitational wave background, similar to electromagnetic black body radiation.
We study the contribution from hidden particles to the diffuse background emitted by the primordial plasma in the early universe.
While this contribution can easily dominate over that from Standard Model particles, we find that both are capped by a generic upper bound that makes them difficult to detect with interferometers in the foreseeable future.
We illustrate our results for axion-like particles and heavy neutral leptons.
Finally, our results suggest that previous works overestimated the gravitational wave background from particle decays out of thermal equilibrium.
\end{abstract}

\newpage
\tableofcontents
\clearpage

\section{Introduction}

Primordial \GWBs are a tantalizing probe of new physics in the early universe
that can complement searches for hints in cosmology or at particle physics experiments.
In particular, the prospective \LISA \cite{LISA:2017pwj} will likely be sensitive to \GW signatures of the strong first-order phase transitions predicted in various popular new physics models
\cite{Caprini:2015zlo,Weir:2017wfa,Caprini:2019egz,LISACosmologyWorkingGroup:2022jok}.
The observation of such backgrounds is a major physics motivation of \LISA and various other proposed \GW detectors, such as the \ET \cite{Punturo:2010zz,Maggiore:2019uih},
\DECIGO \cite{Kawamura:2011zz,Sato:2017dkf} and u-\DECIGO \cite{Seto:2001qf,Kudoh:2005as}, and \BBO \cite{Corbin:2005ny,Crowder:2005nr,Harry:2006fi}.
Recently, significant attention has also turned towards the possibility of detecting ultra high-frequency \GWs, see \eg \cite{Aggarwal:2020olq} and references therein for particular examples of proposed searches.
One advantage of this kind of search is that the known sources of the astrophysical \GWB are not small enough to produce \GWs at high frequencies above $\sim \unit[10^4]{Hz}$  \cite{Aggarwal:2020olq}.

While strong first order phase transitions and other violent out-of-equilibrium phenomena that may have occurred in the early universe
are amongst the most promising sources of primordial \GWBs \cite{Kamionkowski:1993fg,Dasgupta:2022isg,Apreda:2001us,Grojean:2006bp,Caprini:2018mtu,Giovannini:2019oii},
a deviation from thermal equilibrium is not a strict requirement for \GW production:
Any viscous plasma produces \GWs due to microscopic thermal fluctuations \cite{Ghiglieri:2015nfa,Ghiglieri:2020mhm,Ghiglieri:2022rfp}.
The resulting background is generically expected to be much weaker than that of out-of-equilibrium processes,
but it is much harder to avoid since the mechanism does not rely on any deviation from the standard cosmology.
Thermal fluctuations are expected to be a primary source of the \SM
contribution to the primordial \GWB \cite{Ghiglieri:2015nfa,Ghiglieri:2020mhm},
and it has been pointed out that this contribution effectively measures the largest temperature of the early universe plasma after inflation \cite{Ghiglieri:2020mhm,Ringwald:2020ist}.

In general, thermal fluctuations produce a black-body-like \GW spectrum that peaks at momenta $k \sim \pi T$ and exhibits a long tail extending to much smaller frequencies.
Close to the peak, the fluctuations are best described as the result of particle collisions, but this description breaks down for sufficiently small frequencies, where hydrodynamics becomes applicable instead.
In this domain, the production rate exhibits a universal $\propto k^3$ frequency scaling. 
Hydrodynamics is appropriate \eg for computing the \SM  contribution to primordial \GWBs with frequencies in the \LISA window
$\unit[10^{-4}]{Hz} < f < \unit[10^{-1}]{Hz}$ \cite{Ghiglieri:2020mhm,Ringwald:2020ist,LISACosmologyWorkingGroup:2022jok},
and the same can be true for contributions from many extensions of the \SM.
Since the size of the relevant hydrodynamic fluctuations generically scales as the mean free path of the most weakly interacting plasma constituent
\cite{Ghiglieri:2015nfa,Ghiglieri:2020mhm,Hosoya:1983id,Jeon:1994if,Jeon:1995zm},
\GW production can be enhanced if these extensions contain particles whose interactions are weaker than the \SM $\U(1)$ hypercharge gauge interactions.
The existence of such feebly interacting particles \cite{Agrawal:2021dbo} is predicted by various theoretical ideas to address the shortcomings of the \SM,
either as a standalone or as a component of an extended hidden sector that couples to the \SM only via portal interactions, \cf \eg  \cite{Alekhin:2015byh,Curtin:2018mvb,Agrawal:2021dbo,Arina:2021nqi} and references therein. 
These particles can be produced thermally or via non-thermal processes such as (p)reheating or the decay of heavier new particles.
A similar enhancement of \GWBs due to the presence of feebly interacting particles has been pointed out in the context of strong first-order phase-transitions \cite{Jinno:2022fom},
and the production of gravitational waves from hydrodynamic fluctuations has also been considered in the context of axion-like warm inflation \cite{Klose:2022knn,Klose:2022rxh}.

In the present work, we focus on \GWBs produced in thermal equilibrium
during the radiation dominated period of the early universe.
The primary quantity determining the production rate in the hydrodynamic regime is the shear viscosity $\eta$.
The shear viscosities of various generic $\U(N)$ and $\SU(N)$ gauge theories are well established in the literature \cite{Arnold:2000dr,Arnold:2002zm,Arnold:2003zc},
and there also exist prior results for the viscosity of a single, feebly interacting (pseudo-)scalar \cite{Jackson:2018maa,Klose:2022knn}.
We adapt the strategy in \cite{Jackson:2018maa,Klose:2022knn} to compute the shear viscosity of a massive feebly interacting fermion $\psi$,
and combine this result with these prior computations to study the \GWB due to generic feebly interacting particles,
pointing out the existence of a model-independent upper bound.
As specific examples, we then consider \ALPs and \HNLs.

This article is structured as follows:
In \cref{sec:gravitational waves from hydro}, we briefly review the mechanism for thermal \GW 
production.
We present the rates for \GW production from hidden scalars and fermions in \cref{sec:prodrates}, where the latter is based on the computation in \cref{Sec:FermionDiagram}.
In \cref{sec:pheno}, we then discuss the phenomenological implications for \GWBs from feebly interacting particles, 
including a model-independent upper bound (\cref{Sec:UpperBound}) and useful approximate formulae to compute the \GWB in the high temperature regime (\cref{Sec:AnalyticEstimates}).
We illustrate these for two popular hidden particles in section \cref{Sec:Examples}, namely \ALPs and \HNLs. 
\Cref{sec:conclusion} concludes the paper.

\section{Gravitational waves from thermal fluctuations}
\label{sec:gravitational waves from hydro}

\GWs are traceless-transverse tensor perturbations to the metric. 
The relevant line-element for a spatially flat FLRW background cosmology is
\begin{align}
\text{d}s^2 &= \text{d} t^2 - a^2 (\delta_{ij} + h_{ij}) \text{d} x^i \text{d} x^j \ , &
h_i^i &= \partial_i h_{ij} = 0 \ ,
\end{align}
where $a = a(t)$ is the scale-factor of the universe and $h_{ij}$ the graviton field.
It evolves according to the linearized wave-equation
\begin{align}
\ddot h_{ij} + 3 H \dot h_{ij} - \frac{\bm \nabla^2 h_{ij}}{a^2}  &= \frac{16 \pi \, T_{ij}}{a^2 m_\text{Pl}^2} \ ,
\end{align}
where $m_\text{Pl} \equiv \flatfrac1{\sqrt G} \approx \unit[1.22 \cdot 10^{19}]{GeV} $ is the Planck mass,
and $T_{ij}$ the traceless-transverse contribution to the Einstein stress-energy tensor.

In the following, we focus on the production of gravitational waves at sub-horizon scales due to microphysical fluctuations in a thermalized early universe plasma.
In this case, the Hubble rate $H \equiv \flatfrac{\dot a}a$ is small compared to the redshifted \GW frequency $\nicefrac{2\pi f a_0}{a} = k$,
where $f$ is the frequency at present time, $k$ the physical graviton momentum, and $a_0$ the scale factor at present time.
It can be shown that the spectral energy density $e_\text{gw}$ 
evolves according to the equation of motion \cite{Ghiglieri:2015nfa,Ghiglieri:2020mhm,Ringwald:2020ist}
\begin{align}
\label{eq:gravitational wave eom}
\frac{\text{d} \dot e_\text{gw}}{\text{d} \ln f} + 4 H \frac{\text{d} e_\text{gw}}{\text{d} \ln f} &= 16 \pi^2 \left(\frac{f a_0}{a}\right)^3 \frac{\mathrm\Pi(\flatfrac{2\pi f a_0}a)}{m_\text{Pl}^2} \ , &
e_\text{gw} &\equiv \frac{m_\text{Pl}^2}{32\pi} \left\langle \dot h^{ij} (t, \bm 0) \dot h_{ij} (t, \bm 0) \right\rangle_\rho \ ,
\end{align}
where the production rate
\begin{align}
\label{eq:stress energy correlator}
\mathrm\Pi ( k) &= \frac12 \int \hspace{-3pt} \text{d} t  \, \text{d}^3 \bm x \ e^{\i (k t - \bm k \bm x)} \
\mathds L^{ij;kl} \left\langle \left\{ T_{ij} (t, \bm x), T_{kl} (0, \bm 0) \right\} \right\rangle_\rho \ , &
\bm k \bm x &= \delta_{ij} k^i x^j \ , &
k^2 = \delta_{ij} k^i k^j
\end{align}
is computed in locally Minkowskian coordinates.
Accordingly, the spatial Lorentz indices are raised and lowered using the Kronecker delta $\delta_{ij}$, while the projector
\begin{align}
\mathds L_{ij;kl} &= \frac12 \left( \mathds L_{ik} \mathds L_{jl} + \mathds L_{il} \mathds L_{jk} - \mathds L_{ij} \mathds L_{kl} \right) \ , &
\mathds L_{ij} &= \delta_{ij} - \frac{k_i k_j}{k^2} \ ,
\end{align}
ensures that only the traceless-transverse components of $T_{ij}$ contribute.
Working in the Heisenberg picture, the quantum-statistical average $\left\langle \mathcal O \right\rangle_\rho \equiv \tr\left\{ \rho \ \mathcal O \right\}$,
where $\rho$ is the von Neumann density matrix, encodes information about the state of the early universe plasma, including \eg the temperature, at the time of production.

\subsection{Enhancement due to feeble interactions}
\label{Sec:Enhancement}

The stress-energy tensor is a composite operator constructed from the various elementary quantum fields in the theory.
At sufficiently large distances, it is possible to capture the dynamics of systems close to local thermal equilibrium in a model-independent way
by expanding the tensor around its perfect fluid ansatz.
Choosing a frame in which the fluid three-velocity $v^i$ is small, this defines the hydrodynamic stress-energy tensor
\begin{equation}\label{eq:hydrodynamics}
\begin{aligned}
T_0^0 &= e \ , &
T_0^i &= (e+p) v^i \ , &
T_i^j &= \left[ p - \zeta (\bm \nabla \bm v) \right] \delta_i^j - \eta \left[ \partial_i v^j + \partial^j v_i - \frac23 \delta_i^j (\bm \nabla \bm v) \right] \ ,
\end{aligned}\end{equation}
where $e$ is the local energy density, $p$ the pressure, 
$\zeta$ the bulk viscosity, and $\eta$ the shear viscosity.
The hydrodynamic ansatz \eqref{eq:hydrodynamics} is consistent at energy scales $\omega \gg \mathrm\Gamma$,
where $\mathrm\Gamma \sim \nicefrac{k^2 \eta}{T^4}$ is the damping rate of sound waves with wavenumber
$k = \nicefrac{\omega}{c_s}$.
Using that the speed of sound $c_s \approx \nicefrac1{\sqrt3}$ is approximately independent of $k$,
one finds the equivalent condition $\omega \eta \ll T^4$.
In this regime, the shear viscosity $\eta$ measures the size of the traceless-transverse contributions to the stress-energy tensor.
Indeed, the production rate $\mathrm\Pi$ turns out to obey the Kubo formula \cite{Ghiglieri:2015nfa}
\begin{align}
\label{eq:kubo formula}
\lim_{k\to0} \mathrm\Pi = 8 T \eta \ .
\end{align}
The shear viscosity is generically expected to scale as the mean free path $l_\text{av}$
of the most weakly interacting plasma constituent \cite{Hosoya:1983id,Jeon:1994if,Jeon:1995zm},
\begin{align}
\label{eq:shear viscosity}
\eta \sim l_\text{av} v_\text{av} T^4 \sim \nicefrac{T^4}{\mathrm \Upsilon} \ ,
\end{align}
where $v_\text{av}$ is the average velocity associated with the relevant particle species
and $\mathrm\Upsilon$ is its total width, which is connected to the mean free path via the optical theorem. 
In other words, particles with feeble interactions, and therefore long free paths, tend to dominate \GW production in the regime $\omega \eta \ll T^4$,
making \GWs a potentially promising probe for hidden sectors.

At first glance, the scaling of \eqref{eq:shear viscosity} also seems to imply that the shear viscosity,
and with it the \GW production rate, diverges in the limit $\mathrm\Upsilon \to 0$, where the particle is in fact completely sterile.
However, this scaling is only valid in the hydrodynamic limit $k \ll \mathrm\Upsilon$.
For smaller widths $\mathrm\Upsilon \lesssim k$, one instead recovers the naively expected scaling $\mathrm\Pi \propto \mathrm\Upsilon$, with no enhancement for feebly interacting particles.
Hence, as expected, feebly interacting particles dominate primordial \GW production in the hydrodynamic regime but not necessarily in other regimes.

\subsection{Present day spectrum}
\label{sec:present spectrum}

To solve \cref{eq:gravitational wave eom}, one has to fix the time-dependence of the scale factor $a$,
which is connected to the evolution of the early universe plasma via conservation of the comoving entropy density, $\partial_t [a^3 (s + s_h)] = 0$,
where $s$ and $s_h$ respectively denote the \SM and hidden sector contributions to the entropy density.
Neglecting entropy transfer between the two sectors,%
\footnote{This assumption is not always valid.
In particular, heavy new particle decays, \eg in conjunction with a period of early matter domination \cite{Turner:1983he,Ferreira:1997hj,Catena:2009tm}, can inject entropy into the \SM plasma.
The net effect of such an entropy injection would be an additional dilution of primordial \GW signals, which we neglect.}
entropy conservation applies separately to $s$ and $s_h$, so that
\begin{align}\label{eq:entropy conservation}
\partial_t (a^3 s) &= 0 \ , &
s &= \frac{2\pi^2}{45} g_s(T) T^3 \ , &
\frac{a T}{a_0 T_0} &= \left[\frac{g_s(T_0)}{g_s(T)}\right]^{\nicefrac13} \ , 
\end{align}
where $T$ is the temperature of the \SM plasma, $g_s = g_s(T)$ the effective number of radiation degrees of freedom as measured by the \SM entropy density,
and $T_0$ the temperature at present time.
We assume that the \SM dominates the overall energy budget of the universe, so that the temperature evolves according to the standard cosmology,
\begin{align}
\label{eq:temperature evolution}
\frac{\text{d}T}{\text{d}t} &= - \frac{g_s}{g_c} \frac{T^3}{m_0} \ , &
m_0 &\equiv m_\text{Pl} \left[ \frac{4\pi^3 g_\rho}{45} \right]^{-\nicefrac12} \ , &
\rho &= g_\rho \frac{\pi^2}{30} T^4 \ , &
c &= g_c \frac{2\pi^2}{15} T^3 \ ,
\end{align}
where $m_0$ tracks the comoving temperature, and $g_\rho$ and $g_c$ denote the effective number of radiation degrees of freedom as measured by the \SM energy density $\rho$ and heat capacity $c$.
This assumption is reasonable if the total number of hidden sector degrees of freedom is small compared to the number of \SM degrees of freedom.
Combining \cref{eq:gravitational wave eom,eq:entropy conservation,eq:temperature evolution}, the present time stochastic \GWB from thermal fluctuations is \cite{Ghiglieri:2020mhm,Ringwald:2020ist}
\begin{align}\label{eq:gravitational wave spectrum def}
h^2 \Omega_\text{gw}(f)
&\equiv \frac{h^2}{e_\text{crit}} \frac{\text{d} e_\text{gw}}{\text{d} \ln f} 
= \frac{2880 \sqrt{5\pi}}{\pi^2} h^2 \Omega_\gamma \frac{f^3}{T_0^3} \hspace{-3pt} \int\displaylimits_{T_\text{min}}^{T_\text{max}} \hspace{-3pt} \frac{\text{d} T^\prime}{m_\text{Pl}}
\frac{g_c(T^\prime) \left[g_s(T_0) \right]^{\nicefrac13}}{\left[g_\rho(T^\prime) \right]^{\nicefrac12} \left[g_s(T^\prime) \right]^{\nicefrac43}}
\frac{\mathrm\Pi \left( \flatfrac{2\pi f a_0}{a^\prime} \right)}{8T^{\prime 4}} \ ,
\end{align}
where
\begin{align}
h^2 \Omega_\gamma &= \frac{h^2}{e_\text{crit}} \frac{2\rho}{g_\rho(T_0)} \ , &
e_\text{crit} &= \frac{3 H_0^2 m_\text{Pl}^2}{8\pi} \ , &
H_0 &= \unit[100]{km \, s^{-1} Mpc^{-1}} \cdot h
\end{align}
is the present day photon energy density, normalized to the critical energy density $e_\text{crit}$, $H_0$ the present day Hubble rate, and $h$ the reduced Hubble rate.
Assuming that the bulk of the \GWB is produced at temperatures well above the electroweak phase transition, one has
$g_c(T^\prime) = g_s(T^\prime) = g_\rho(T^\prime) = 106.75$.
Also using \cite{Fixsen:2009ug,Saikawa:2018rcs,ParticleDataGroup:2022pth}
\begin{align}
\label{eq:photon energy density numbers}
h^2 \Omega_\gamma &= 2.4728(21) \cdot 10^{-5} \ , &
g_s(T_0) &= 3.931(4) \ , &
T_0 &= \unit[2.7255(6)]{K} = \unit[3.5682(7)\cdot10^{11}]{Hz} \ ,
\end{align}
one obtains\footnote{%
The size of the final \GWB can be expressed also in terms of the characteristic strain $h_c$, which is defined as
\begin{align}\label{eq:gw strain}
h_c^2 = \frac{3H_0^2}{2\pi^2f^2}\Omega_\text{gw} \propto \frac{h^2 \Omega_\text{gw}}{f^2} \ .
\end{align}
}
\begin{align}\label{eq:gravitational wave spectrum}
h^2 \Omega_\text{gw}(f)
&\approx 
2.02 \cdot 10^{-38} \times \left( \frac{f}{\unit{Hz}} \right)^3
\times \hspace{-3pt} \int\displaylimits_{T_\text{min}}^{T_\text{max}} \hspace{-3pt} \frac{\text{d} T^\prime}{m_\text{Pl}} \frac{\mathrm\Pi \left( \flatfrac{2\pi f a_0}{a^\prime} \right)}{8T^{\prime 4}} \ .
\end{align}
This is the main expression we use in the remainder of this work.
If the production rate $\mathrm\Pi$ and the integration bounds $T_\text{max}$ and $T_\text{min}$ are independent of $f$,
the prefactor results in the characteristic $\propto f^3$ frequency shape associated with backgrounds from hydrodynamic fluctuations. 
This has also been observed \eg within the context of \GWs from
post-inflationary phases stiffer than radiation \cite{Giovannini:1998bp} and axion-like inflation \cite{Klose:2022rxh}.

\Cref{eq:gravitational wave spectrum def} explicitly relies on our assumptions that the \SM is in kinetic equilibrium and dominates the overall energy budget,
while expression \eqref{eq:gravitational wave spectrum} further assumes that its number of radiation degrees of freedom also evolves according to standard cosmic history.
This implies that expression \eqref{eq:gravitational wave spectrum}, and subsequently our results in \cref{sec:pheno}, only applies to \GWBs produced after inflation and
after the plasma of inflaton decay products has equilibrated \cf~\eg~\cite{Mukaida:2015ria}. 
\footnote{See appendix \ref{sec:non standard temperature evolution appendix} for formulae equivalent to \eqref{eq:gravitational wave spectrum def}
and \eqref{eq:gravitational wave spectrum} for more general expansion histories that \eg deviate from radiation domination.}
While \GWBs from earlier epochs can be important, and although, in principle, the present formalism can be used to compute them,
we stick to \eqref{eq:gravitational wave spectrum def} and only briefly comment on \GW production during (p)reheating at the end.
That being said, \cref{eq:gravitational wave spectrum def,eq:gravitational wave spectrum} do not necessarily require the hidden sector to be in thermal equilibrium.
In fact, they contain no assumption about the hidden sector phase space distribution functions except that their total energy density has to be small compared to the \SM one.
The distribution functions only enter at the level of computing $\mathrm\Pi \left( \flatfrac{2\pi f a_0}{a} \right)$. 
For practical purposes, we consider equilibrium distributions in most of section \ref{sec:prodrates}
(\cf \eqref{eq:gw_scalar_prodrate} and \eqref{eq:fermionic prodrate}), but we emphasize that this assumption is not crucial,
and that the generalisation of our results to non-equilibrium situations is straightforward (\cf \eqref{GeneralPi}). 
In sections \ref{sec:pheno} and \ref{Sec:Examples}, we for simplicity further assume that the \SM and hidden sector temperatures are equal to each other.\footnote{%
To satisfy observational constraints \eg from big bang nucleosynthesis, thermalized hidden sectors are generically expected to be colder than the \SM plasma \cite{Breitbach:2018ddu,Freese:2023fcr},
but it has been argued that larger temperatures can be viable in certain cases \cite{Tenkanen:2016jic,Ertas:2021xeh}.
}
If the hidden sector temperature deviates from that of the \SM, our assumptions imply that their ratio is approximately a model-dependent constant over the domain of integration in \cref{eq:gravitational wave spectrum}.
Hence, the discussion can be generalized to other hidden sector temperatures by an appropriate re-scaling of the production rate $\mathrm\Pi$. 
Finally, in practice, we consider the contribution from one hidden particle at a time. If there are several hidden particles, the signals are additive in the enhanced regime as long as the total energy density is dominated by the \SM.

\section{Gravitational wave production rates}
\label{sec:prodrates}

The \GW production rate $\mathrm\Pi$ is the final ingredient needed to predict the \GWB \eqref{eq:gravitational wave spectrum}.
In this section, we first summarize some existing results for thermal equilibrium production due to the \SM and a generic feebly interacting (pseudo-)scalar,
and then present the contribution due to a generic feebly interacting fermion.
We finally provide compact summary formulae that encompass both the scalar and fermionic cases and are suitable for phenomenological applications.

In the \LISA frequency window, and for sufficiently large maximal temperatures,
the \SM contribution is dominated by hydrodynamic fluctuations of the right-handed leptons \cite{Arnold:2000dr,Arnold:2002zm,Arnold:2003zc},
yielding the production rate \cite{Ghiglieri:2015nfa}
\begin{align}\label{eq:sm gw production rate}
\mathrm\Pi^\text{SM} = 8 \eta_{\SM} T = \frac{8\pi^2 g_{\SM}}{225} \frac{T^5}{\mathrm\Upsilon_{\SM}} &\approx \frac{16 T^4}{g^4 \ln( \frac{5 T}{m_D})} \approx 400 T^4 \ , &
m_D^2 &= \frac{11}{6} g^2 T^2 \ ,
\end{align}
where $g \approx 0.36$ is the $\U(1)$ hypercharge gauge coupling, $m_D$ the associated Debye mass, and $g_{\SM} = 6$ the degrees of freedom associated with the right-handed leptons.
The width $\mathrm\Upsilon_{\SM}$ is defined consistent with our conventions in \eqref{eq:relat rate},
which simplifies comparing the \SM and new physics contributions.
Ignoring the running of the weak hypercharge coupling, \eqref{eq:sm gw production rate} yields energy density
\begin{align}
\label{eq:final gw background}
h^2 \Omega_\text{gw}^\text{SM}(f)
&\simeq 1.03 \cdot 10^{-36} \times \left( \frac{f}{\unit{Hz}} \right)^3 \times \frac{T_\text{max}}{m_\text{Pl}} \ ,
\end{align}
which depends only on $T_\text{max}$.
For modes that never leave the horizon, this temperature is closely related to the reheating temperature after inflation, and can probe the dynamics of reheating \cite{Ghiglieri:2020mhm,Ringwald:2020ist}.
Given a model of inflation, 
the CMB is sensitive to the reheating epoch \cite{Liddle:2003as,Martin:2010kz,Adshead:2010mc}, 
and next-generation observatories can potentially independently determine the reheating temperature \cite{Drewes:2022nhu,Drewes:2023bbs},
highlighting the complementarity of multi-messenger probes of the early universe.

\subsection{Feebly interacting particles}
\label{sec:shear viscosity}

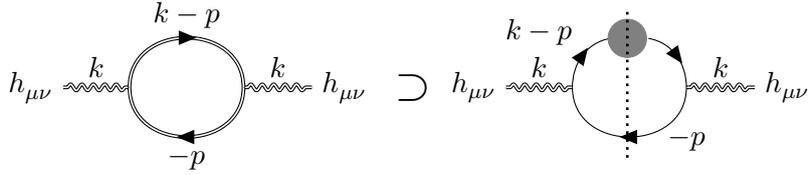
\begin{figure}
\begin{tikzpicture}
  \begin{feynman}[every blob={/tikz/fill=gray!30,/tikz/inner sep=2pt}]
    \vertex (a) {$h_{\mu\nu}$};
    \vertex [right=of a,xshift=-.2cm] (b);
    \vertex [right=of b] (c);
    \vertex [right=of c,xshift=-.6cm] (d) {$h_{\mu\nu}$};

    \diagram* {
      (a) -- [graviton,edge label=$k$] (b),
      (c) -- [graviton, edge label=$k$] (d),
      (c) -- [doublefermion, half left, edge label=$-p$] (b),
      (b) -- [doublefermion, half left, edge label=$k-p$] (c),
    };
    \vertex [right=of d,xshift=-.6cm] (equalnode) {\Large $\supset$ \normalsize};

    \vertex [right=of d,xshift=.2cm] (E) {$h_{\mu\nu}$};
    \vertex [right=of E,xshift=-.2cm] (F);
    \vertex [right=of F] (G);
    \vertex [right=of G,xshift=-.6cm] (H) {$h_{\mu\nu}$};
    \vertex[right=of F,xshift=-1cm,yshift=.65cm] (internalvertex1);
    \vertex[right=of F,xshift=-.5cm,yshift=.65cm] (internalvertex2);
    \vertex[below=of G,yshift=1.1cm] (labelvertex) {$-p$};
    \diagram* {
      (E) -- [graviton,edge label=$k$] (F),
      (G) -- [graviton, edge label=$k$] (H),
      (G) -- [fermion, half left] (F),
      (F) -- [fermion, in=190,out=90, edge label=$k-p$] (internalvertex1),
      (internalvertex2) -- [fermion, in=90,out=-10] (G),
    };
    \filldraw[right=of d,xshift=7.85cm,yshift=.65cm,gray,opacity=1] circle (7.2pt);
    \draw[dotted,line width= 1pt] (7.85,1) to (7.85,-1);
  \end{feynman}
\end{tikzpicture}
\caption{\label{fig:wave_diag}
To find the \GW production due to a feebly interacting fermion, we compute the resummed Feynman diagram shown on the left-hand side, where the straight double lines represent resummed $\psi$ propagators.
Our result encompasses an infinite class of diagrams in fixed-order perturbation theory, including the one depicted on the right-hand side, where the grey blob denotes a single self-energy insertion.
The dotted line signaling cut propagators, the optical theorem implies that it and other diagrams we include together encode the rate of producing gravitons as bremsstrahlung.
}
\end{figure}

The production rate due to hydrodynamic fluctuations of a single feebly interacting (pseudo-)scalar is \cite{Jackson:2018maa,Klose:2022knn}%
\footnote{
The complete expression for the function $F(x,v)$ is
\begin{multline}
\label{eq:angular function}
F(x,v) = \frac{30(4 x^2 + 5 x^2 (1-v^2) - 3)}{48 v^4 x^4} + \frac{30x(1-x^2(1-v^2))}{16 v^5 x^5} \ln \frac{1+x^2 (1+v)^2}{1+x^2 (1-v)^2} \\
+ \frac{15(1-x^2(1-v^2))^2 - 60 x^2}{16 v^5 x^5} \arctan \frac{2v x}{1+x^2(1-v^2)} \ .
\end{multline}}
\begin{align}
\label{eq:gw_scalar_prodrate}
\mathrm\Pi &\simeq
\int \text{d} p \, p^6 \frac{4 \, n_B(\epsilon) (1+n_B(\epsilon))}{15 \pi^2 \epsilon^2} \frac1{2 \mathrm\Upsilon_p} F\left( \frac{k}{2 \mathrm\Upsilon_p}, \frac{p}{\epsilon} \right) \ , &
F(x,v) &= 
\begin{cases}
\frac1{1+x^2}  & v \ll 1 \\
1 & v = 1 \land x \ll 1 \\
\frac5{2 x^2} & v = 1 \land x \gg 1
\end{cases}
\ ,
\end{align}
where $\mathrm\Upsilon_p$ is the finite temperature width of the new particle, $\epsilon^2 = \bm p^2 + m^2$ is its energy squared,
and $n_B(\epsilon) = [\exp(\beta \epsilon) - 1]^{-1} $ is the usual Bose-Einstein distribution. 
The  corresponding expression for fermions is dominated by the resummed diagram shown in figure~\ref{fig:wave_diag};
it coincides with \eqref{eq:gw_scalar_prodrate} up to an overall factor $2 c_X$ that counts the internal degrees of freedom ($c_X = 1$ for a Majorana fermion and $c_X = 2$ for a Dirac fermion),
\begin{align}\label{eq:fermionic prodrate}
\mathrm\Pi(k) &\overset{k \ll \alpha^2 T}{\simeq} 2 c_X \int\displaylimits_0^\infty \text{d} p \, p^6
\frac{4\, n_F(\epsilon) (1- n_F(\epsilon))}{15 \pi^2 \epsilon^2} \frac1{2 \mathrm\Upsilon_p} F\left( \frac{k}{2 \mathrm\Upsilon_p}, \frac{p}{\epsilon} \right) \ ,
\end{align}
where $F(x,v)$ is defined as in expression \eqref{eq:angular function} 
and $n_F(\epsilon) = \left[\exp(\beta \epsilon) + 1 \right]^{-1}$ is the Fermi-Dirac distribution.
Although our primary focus is the enhanced regime $x\ll 1$, where $F(x,v) \to 1$, we note that expression \eqref{eq:fermionic prodrate} is also a valid approximation
of the wave-function type diagram in \cref{fig:wave_diag} for larger frequencies $\mathrm\Upsilon_p < k \ll T$.
As indicated by the Bose-Einstein and Fermi-Dirac distributions $n_B(p_0)$ and $n_F(p_0)$, the expressions \eqref{eq:gw_scalar_prodrate} and \eqref{eq:fermionic prodrate} assume that the hidden sector is in thermal equilibrium.
They can be generalised to non-equilibrium expressions in which the distributions are replaced by general phase space distribution functions $n(p_0)$.
We present the derivation for the fermionic case in appendix \ref{Sec:FermionDiagram}, finding the result
\begin{align}\label{GeneralPi}
\mathrm\Pi(k) &\simeq \frac{c_X}2 \int \frac{\text{d}^3 \bm p}{(2\pi)^3} p_\perp^4  \left[D (\epsilon, \epsilon) + D(-\epsilon,-\epsilon) \right]
\frac{2 \mathrm\Upsilon_p} {k^2 (\epsilon - p_\sparallel)^2 + 4 \epsilon^2 \mathrm\Upsilon_p^2} \ ,
\end{align}
where
\begin{align}\label{Ddef}
D(p_0, q_0) &= (1-n(p_0)) n(q_0) + n(p_0) (1 - n(q_0)) \ .
\end{align}
In thermal equilibrium, one has $D(\epsilon,\epsilon) = 2 n_F(\epsilon) (1-n_F(\epsilon))$.
Inserting this expression and evaluating the angular integral, we obtain \eqref{eq:fermionic prodrate} as a limit of \eqref{GeneralPi}.

\subsection{Approximate formulae for phenomenology}
\label{sec:master formula}

The production rate \eqref{eq:fermionic prodrate} is remarkably similar to its (pseudo-)scalar equivalent \eqref{eq:gw_scalar_prodrate}.
In the high ($m \ll T$) and low temperature ($m \gg T$) cases, we can combine the two expressions into a single, approximate formula
by replacing the momentum-dependent width $\mathrm\Upsilon_p$ with its thermal average
\begin{align}
\mathrm \Upsilon_\text{av} &\equiv \frac1{2\pi^2 N_\text{eq}} \int\displaylimits_0^{\infty} \text d p \ p^2 n_X (\epsilon) \mathrm\Upsilon_p \ , &
N_\text{eq} &= \frac1{2\pi^2} \int\displaylimits_0^{\infty} \text d p \ p^2 n_X (\epsilon) \ ,
\end{align}
where $n_X \in \left\{ n_B, n_F \right\}$ is either the Bose-Einstein or Fermi-Dirac distribution, depending on the spin of the hidden particle.
Using $\mathrm\Upsilon_\text{av}$ and setting either $v = \nicefrac{p}{\epsilon} \to 0$ or $v = \nicefrac{p}{\epsilon} \to 1$,
the function $F(x,v)$ becomes momentum-independent and can be pulled out of the integrals in eqs. \eqref{eq:gw_scalar_prodrate,eq:fermionic prodrate}.
In the low-temperature case, it takes on the very simple form
\begin{align}
	F(x,v) \overset{v\to0}{=} \frac1{1+x^2} \ ,
\end{align}
but the high-temperature case is more complicated.
For phenomenological applications, we can nevertheless approximate $F(x,v)$ by interpolating the function between $x \ll1$ and $x\gg1$,
which gives
\begin{align}
	F(x,v) \overset{v\to1}{\approx} \frac5{5+ 2 x^2} \ .
\end{align}
We expect this approximation to be worst for $x \approx 1$.
Indeed, setting $x=v=1$, the above expression yields $F(x,v) \approx 0.71$, while the exact expression \eqref{eq:angular function} yields $F(x,v) \approx 0.56$,
so that the approximation overestimates the gravitational wave production rate by roughly $30 \%$.
However, this is still sufficient for an order of magnitude estimate of the gravitational wave background.
In addition, the extremal regimes $x \ll1$ and $x\gg1$ are generically expected to be the most relevant for phenomenological applications.

Using $\mathrm\Upsilon_\text{av}$, and pulling out $F(x,v)$, the remaining integrals in \eqref{eq:fermionic prodrate,eq:gw_scalar_prodrate} can be computed analytically.
For high temperatures, one obtains
\begin{align}\label{eq:relat rate}
	\mathrm\Pi(k) &\overset{m \ll T}{\simeq} g_X \frac{16 \pi^2}{225} T^5 \frac{5 \mathrm\Upsilon_\text{av}}{k^2 + 10 \mathrm\Upsilon_\text{av}^2} \ , &
	g_X &=
	\begin{cases}
		1 & \text{Spin $0$} \\
		2 \cdot c_X & \text{Spin $\nicefrac12$}
	\end{cases} \ ,
\end{align}
where we have neglected a relative factor of $\nicefrac78$ in fermionic case that originates from the distribution functions. 

For low temperatures, one has to distinguish between particles that retain a relativistic momentum distribution,
\eg because they undergo relativistic freeze-out at some temperature $T_f \gg m$, and particles whose distribution function becomes non-relativistic.
For the relativistic case, one obtains 
\begin{align}\label{eq:relat freeze}
	\mathrm\Pi(k) &\overset{T \lesssim m \ll T_f}{\simeq} g_X \frac{64 \pi^4}{315} \frac{T^7}{m^2} \frac{2 \mathrm\Upsilon_\text{av}}{k^2 + 4 \mathrm\Upsilon_\text{av}^2} \ ,
\end{align}
where we have now neglected a relative factor of $\nicefrac{31}{32}$ in the fermionic case.
Note that this rate is independent of the freeze-out temperature.
For particles that assume a non-relativistic momentum distribution, freezing out at some temperature $T_f \lesssim m$, one obtains the expression
\begin{align}\label{eq:nonrelat freeze}
	\mathrm\Pi(k) &\overset{T, T_f \lesssim m}{\simeq}
	g_X \left( \frac{2}{\pi} \right)^{\nicefrac32} \left[ \frac{T^7}{m^2} \left( \frac{\xi m}{T} \right)^{\nicefrac72} e^{-\nicefrac{\xi m}{T}} \right] \frac{2 \mathrm\Upsilon_\text{av}}{k^2 + 4 \mathrm\Upsilon_\text{av}^2} \ , &
	\xi &=
	\begin{cases}
		1 & T > T_f \\
		\nicefrac{T}{T_f} &T < T_f 
	\end{cases} \ ,
\end{align}
which explicitly depends on $T_f$.
Setting $T_f \to 0$, this expression also captures particles that never freeze-out but always remain in thermal equilibrium.
This is the case \eg for unstable particles with a large width $\mathrm\Upsilon_\text{av} > H$, which remain in thermal equilibrium as they decay.
In contrast, we expect that unstable particles with a \emph{small} width $\mathrm\Upsilon_\text{av} \ll H$ are better captured by the relativistic formula \eqref{eq:relat freeze},
because the explicit $\propto \nicefrac{T^7}{k^2} \sim T^5$ suppression in \eqref{eq:relat freeze} effectively cuts off \GW production well before decays can significantly impact their momentum distribution and number density.

Finally, we note that the rate for a relativistic distribution is strictly larger than the one for a non-relativistic distribution,
so that it can be used to obtain a model-independent upper bound on \GW production, \cf~section \ref{Sec:UpperBound}.

\section{Phenomenology}
\label{sec:pheno}

In this section, we apply the results of \cref{sec:prodrates} to study the phenomenology and the prospects of detecting \GWBs produced by feebly interacting particles.
The full background can be computed by numerically integrating the production rates in~\eqref{eq:gw_scalar_prodrate} and~\eqref{eq:fermionic prodrate}.
To help illustrate its qualitative features, we also derive analytical expressions that are applicable in various \GW production regimes.

\subsection{Generic features}

The \GW production rate is proportional to $k^3 \mathrm\Pi(k)$.
In thermal equilibrium, it depends on the physical graviton momentum $k =\nicefrac{2\pi f a_0}{a}$,
the temperature of the early universe plasma $T$, and the properties of the particles driving production.
The \SM contribution \eqref{eq:sm gw production rate} is determined by the width $\mathrm\Upsilon_{\SM}$,
which measures the free-streaming length of the right-handed \SM leptons.
Likewise, the new physics contributions \eqref{eq:gw_scalar_prodrate,eq:fermionic prodrate} due a single feebly interacting particle
mainly depend on its mass $m$ and the width $\mathrm\Upsilon_\text{av}$, which characterizes the free-streaming length of the new particle.

The new physics contribution dominates the \SM contribution in the regime $k < \mathrm\Upsilon_\text{av} \ll \mathrm\Upsilon_{\SM}$.
In this regime, hydrodynamics governs both sectors.
The new physics contribution \eqref{eq:fermionic prodrate} scales as $\nicefrac{k^3}{\mathrm\Upsilon_\text{av}}$
while the smaller \SM contribution \eqref{eq:sm gw production rate} scales as $\nicefrac{k^3}{\mathrm\Upsilon_{\SM}}$,
so that both contributions grow like $k^3$.
For larger momenta $\mathrm\Upsilon_\text{av} < k < \mathrm\Upsilon_{\SM}$,
hydrodynamics still governs the \SM but not the hidden sector.
It is no longer necessary to resum wave-function type corrections to the hidden particle propagators,
so that a more Boltzmann-like picture of the hidden sector as a particle gas becomes appropriate.
The new physics contribution now scales as $k \mathrm\Upsilon_\text{av}$, growing like $k$.
Since the \SM contribution continues to grow like $k^3$, it eventually overtakes the hidden sector to dominate the overall production rate.
For even larger momenta $\mathrm\Upsilon_{\SM} < k$, the Boltzmann picture captures both the \SM and the hidden sector.
The overall production rate still grows and finally peaks at momenta $k \sim \pi T$.
Beyond this peak, the production rate is Boltzmann-suppressed 
and therefore effectively cut off.

The most important parameter determining the magnitude of the final background is the maximal temperature $T_\star$ of the
hidden sector plasma, which is a highly model dependent and encodes \eg the specific dynamics of reheating.
Although we assume that the \SM and hidden sector plasmas share the same temperature $T$,
it is possible for the \SM to thermalize well before or after the hidden sector, leading to different effective values of $T_\star$ for the two contributions.
Along this line of reasoning, one might worry that it is impossible to produce a sufficient quantity
of thermally distributed feebly interacting hidden particles early enough to result in a significant contribution in the final \GWB.
However, if the \SM is produced before the hidden sector, the \SM can produce thermalized hidden particles via freeze-in, which puts an effective lower bound on $T_\star$.
Explicitly, interactions between the \SM and hidden sectors equilibrate for temperatures $\mathrm\Upsilon_\text{av} \sim H \propto \nicefrac{T^2}{m_\text{Pl}}$. 
One way to evade this worst-case scenario, which can still result in a significant new physics contribution to the final \GWB,
is to produce the hidden sector either via non-renormalizeable interactions (\cf \cref{sec:nonrenorm interactions}) or non-thermally, \eg via (p)reheating or the decay of heavy new particles.

Focusing on a fixed frequency $f$, the production rate changes over time as the temperature of the early universe plasma decreases.
Since the physical momentum $k \propto \nicefrac{f}{a}$ redshifts the same as the temperature $T\propto \nicefrac1{a}$,
the ratio $\nicefrac{k}{T}$ remains approximately constant,
and the peak of the production rate at $k\sim \pi T$ translates into the same temperature-independent frequency
\begin{align}
f_\text{peak} &\equiv \frac{T_0}2 \left[ \frac{g_s(T_0)}{g_s(T)} \right]^{\nicefrac13} \approx \unit[6 \cdot 10^{10}]{Hz} \ ,
\end{align}
which approximately determines the peak frequency of the final \GWB.
In the following, our primary focus is on the phenomenology of direct detection experiments, which are typically sensitive to much lower frequencies with redshifted momenta $k \ll T$.
These modes can be outside the causal horizon ($k < H$) at the onset of \GW production.
Since super-horizon modes of tensor perturbations are static (\cf \eg~section 5.4 in \cite{Weinberg:2008zzc}), 
production of these modes is delayed until they re-enter the horizon.
Adopting the same standard cosmology as in \cref{sec:present spectrum}, a given mode with frequency $f$ re-enters the horizon at the entry temperature
\begin{align}
T_\text{entry} (f) &= m_0 \frac{\pi f}{f_\text{peak}} \ , &
H(T = T_\text{entry}) &= k \ .
\end{align}
Modes with $T_\text{entry} > T_\star$ enter the horizon before \GW production begins, causing no delay,
while modes with $T_\text{entry} < T_\star$ start to be produced only upon re-entry,
so that one has to replace $T_\star \to T_\text{entry}$ in the integral \eqref{eq:gravitational wave spectrum}.
Hence, the maximal temperature $T_\text{max}$ is
\begin{align}\label{eq:Tmax interp}
T_\text{max} &= \min\left( T_\text{entry}, T_\star \right)
= \frac{\min\left( f, f_\star \right)}{f_\star} T_\star 
\simeq \frac{f}{f + f_\star} T_\star \ ,
\end{align}
where the horizon frequency $f_\star$ is the largest frequency for which production is delayed by the causal horizon,
\begin{align}\label{eq:fstar def}
f_\star &= \frac{f_\text{peak}}{\pi} \frac{T_\star}{m_0} \approx \unit[3\cdot 10^{11}]{Hz} \frac{T_\star}{m_\text{Pl}} \ , &
T_\text{entry}(f_\star) &= T_\star \ .
\end{align}
For sufficiently small frequencies, the entry temperature can drop below the mass of the hidden particle, $T_\text{entry} < m$,
so that \GW production is restricted to the non-relativistic regime.
The frequency at which this happens is the non-relativistic frequency
\begin{align}\label{eq:cutoff frequency}
f_\text{nr} &= \frac{f_\text{peak}}{\pi} \frac{m}{m_0} \approx \unit[3\cdot 10^{11}]{Hz} \frac{m}{m_\text{Pl}} \ , &
T_\text{entry} (f_\text{nr}) &= m \ .
\end{align}
In principle, the entry temperature can even drop below the minimal temperature $T_\text{min}$.
However, this effect is less relevant, since we generically expect to be able to consider minimal temperatures \eg in the $\unit{TeV}$ range or even lower. 
In this case, the frequency for which $T_\text{entry}$ drops below $T_\text{min}$ is less than $\unit[10^{-6}]{Hz}$ and therefore well below the regime of phenomenological interest.

In the regime $f \ll f_\text{peak}$ (or equivalently $k \ll T$), \GW production is governed by the phenomenological formulae we derived in \cref{sec:master formula}.
At high temperatures $T \gg m$, the new physics contribution to the relevant production rate is independent of the mass $m$ and can be approximated using the phenomenological expression \eqref{eq:relat rate}.
Once the temperature falls below the mass, $T \lesssim m$, one has to distinguish between particles that retain their relativistic momentum distribution,
\eg due to relativistic freeze-out, and particles which assume a non-relativistic distribution.
If the hidden particle remains in thermal equilibrium with the \SM plasma, its contribution to the production rate becomes Boltzmann suppressed and becomes effectively irrelevant.
However, if the particle freezes-out or if it is unstable but long lived, with a width $\mathrm\Upsilon_\text{av} \ll H$, \GW production can continue in the low temperature regime $T < m$ for some time.
The phenomenological expression \eqref{eq:relat freeze} captures low-temperature production in case of a relativistic freeze-out,
while expression \eqref{eq:nonrelat freeze} captures the case of a non-relativistic freeze-out.

\subsection{Model-independent upper bound}\label{Sec:UpperBound}

The phenomenological formulae \eqref{eq:relat rate,eq:relat freeze} imply an upper bound on the \GWB for frequencies $f \ll f_\text{peak}$
that is saturated at the transition between the hydrodynamic and Boltzmann regimes.
Explicitly, the high- and low temperature expressions \eqref{eq:relat freeze,eq:relat rate} are maximal for widths
\begin{align}\label{eq:maximal momentum}
\mathrm\Upsilon_\text{av} &=
\begin{cases}
\nicefrac{k}{\sqrt{10}} & T > m \\
\nicefrac{k}{2} & T < m
\end{cases} \ .
\end{align}
Inserting these values into the formulae \eqref{eq:relat rate,eq:relat freeze}, one obtains two separate bounds on the high-temperature contribution,
\begin{align}\label{eq:high temp max}
h^2 \Omega_\text{gw}^{T>m}(f)
&< 2.6 \cdot 10^{-29} \times \left( \frac{f}{\unit{Hz}} \right)^2
\times g_X \frac{\max(T_\text{max}, m) - \max(T_\text{min}, m)}{m_\text{Pl}} \ ,
\end{align}
and the low temperature contribution,
\begin{align}\label{eq:low temp max}
h^2 \Omega_\text{gw}^{T<m}(f)
&< 1.6 \cdot 10^{-28} \times \left( \frac{f}{\unit{Hz}} \right)^2
\times g_X \frac{\min(T_\text{max}^3, m^3) - \min(T_\text{min}^3, m^3)}{m^2 m_\text{Pl}} \ ,
\end{align}
to the final gravitational wave energy density. 
Maximizing the sum of both contributions, we set $T_\text{min} \to 0$ and $m \to T_\text{max}$.
Inserting expression \eqref{eq:Tmax interp}, this gives
\begin{align}\label{eq:simplified max}
h^2 \Omega_\text{gw}(f) &< 4.9 \cdot 10^{-40}
\times g_X \left( \frac{f}{\unit{Hz}} \right)^2 
\min\left(\frac{f}{\unit{Hz}}, \frac{f_\star}{\unit{Hz}} \right) \ .
\end{align}
This expression depends only on $g_X$ and $T_\star$.
We can eliminate the $T_\star$ dependence by using that $\min(f, f_\star) < f$, finding the slightly weaker bound
\begin{align}\label{eq:f small max}
h^2 \Omega_\text{gw}(f)
&< 4.9 \cdot 10^{-40} \times g_X \left( \frac{f}{\unit{Hz}} \right)^3\ ,
\end{align}
where $g_X$ is now the only remaining model-dependent parameter.
On the other hand, considering feebly interacting particles with a given mass $m$, and focusing on frequencies $f < f_\text{nr}$,
so that \GW production is delayed until the particle is non-relativistic, we obtain the much more stringent bound
\begin{align}\label{eq: fixed mass bound}
h^2 \Omega_\text{gw}(f)
&\overset{f<f_\text{nr}}{<} 4.9 \cdot 10^{-40} \times g_X \left( \frac{f}{\unit{Hz}} \right)^3 \frac{f^2}{f^2_\text{nr}} \ .
\end{align}
The $h^2 \Omega_\text{gw} \propto f^5$ scaling of this bound shows that the final \GW spectrum is effectively cut off below $f = f_\text{nr}$.
If we require \eg $f_\text{nr} < \unit[10^{-1}]{Hz}$, such that the entire region of phenomenological interest lies above this cut-off,
one finds that the mass of the feebly interacting particle responsible for \GW production should be $m < \unit[1.1 \cdot 10^8]{GeV}$.

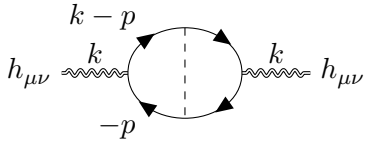
\begin{figure}
\begin{tikzpicture}
  \begin{feynman}
    \vertex (a) {$h_{\mu\nu}$};
    \vertex [right=of a,xshift=-.2cm] (b);
    \vertex [right=of b] (c);
    \vertex [right=of c,xshift=-.6cm] (d) {$h_{\mu\nu}$};
    \vertex [right=of b,xshift=-.75cm,yshift=.6cm] (internalvertex1);     
    \vertex [right=of b,xshift=-.75cm,yshift=-.6cm] (internalvertex2);     

    \diagram* {
      (a) -- [graviton,edge label=$k$] (b),
      (c) -- [graviton, edge label=$k$] (d),
      (b) -- [fermion, out=90, in=180, edge label=$k-p$] (internalvertex1),
      (internalvertex1) -- [fermion, out=0, in=90] (c),
      (internalvertex2) -- [fermion, in=-90, out=180,edge label=$-p$] (b),
      (c) -- [fermion, in=0, out=-90] (internalvertex2),
      (internalvertex1) -- [scalar] (internalvertex2),
    };
  \end{feynman}
\end{tikzpicture}
\caption{\label{fig:vertex_diag} Example of a vertex-type correction to the rate of \GW production from feebly interacting fermions.}
\end{figure}

These bounds, of course, rely on a number of implicit assumptions.
For example, we have not included contributions to the production rate generated by vertex-type diagrams such as the one shown in \cref{fig:vertex_diag} or higher order diagrams.
However, we expect that including such diagrams will at most yield an order one correction to the overall \GW production rate and should not significantly modify the bounds. 
We also assumed that no entropy is transferred from the \SM to the hidden sector.
If the \SM plasma can dump a significant amount of entropy into the hidden sector, this would cause the \GWB to be diluted less than in standard cosmology and therefore enhance it.
Of course, it is also possible to transfer entropy from the hidden sector to the \SM plasma, but this does not impact the validity of our bounds, since it would simply enhance the dilution of the \GWB.

Finally, the bounds \eqref{eq:high temp max,eq:low temp max} only apply to \GWs emitted by thermal equilibrium fluctuations of the early universe plasma. 
They explicitly do not apply to \GWBs from out-of-equilibrium processes with very different underlying physics, such as strong first order phase transitions,
production during (cold) inflation, or  turbulence during preheating (\cf~\cite{Caprini:2018mtu}).
For example, a number of recent works have found that particle decays during reheating can produce a background
significantly in excess of these upper bounds \cite{Kanemura:2023pnv,Barman:2023rpg,Barman:2023ymn,Bernal:2023wus}. 
However, these works seem to compute the \GW production rate without accounting for the resummation of wave-function type corrections that is necessary in the hydrodynamic regime.
We expect that systematically incorporating these effects, using \eg the approach we have employed in \cref{sec:prodrates},
which results in an expression for the production rate \eqref{eq:fermionic prodrate} that can already accommodate out-of-equilibrium distribution functions,
would yield a significant suppression of the resulting \GWB.

\subsection{Analytic estimates of the high-temperature background}\label{Sec:AnalyticEstimates}

The width $\mathrm\Upsilon_\text{av}$ determines the more detailed features of the spectrum.
It generically becomes smaller as the temperatures decreases and depends on the mass of the hidden particle $m$, the temperature $T$,
any number of coupling constants, and potentially some heavy new physics scale $\Lambda \gg m, T$.
If the width is dominated by interactions with the same mass dimension $d$, and neglecting the temperature dependence of running couplings,
it is expected to scale as 
\begin{align}
	\label{eq:width scaling}
	\mathrm \Upsilon_\text{av} &\simeq y \, T \left( \frac{T}{\Lambda} \right)^{2(d-4)} 
	\begin{cases}
		1 & T \gg m \\
		\left( \nicefrac{m}{T} \right)^{n} & T \lesssim m 
	\end{cases} \ , &
	n&\leq 1 + 2(d-4) \ ,
\end{align}
where $y$ is some combination of coupling constants and other dimensionless parameters.
The integer $n$ depends on the properties of the hidden particle and its interactions.
For an unstable particle, the width becomes temperature independent for $T\to0$, so that $n = 1 + 2(d-4)$,
but if the particle is stable $n$ is a model-dependent parameter.
If the low-temperature width is dominated by $2\leftrightarrow2$ scattering events involving relativistic particles in the plasma,
it is related to the thermal average of the total scattering cross section \cite{Kolb:1990vq},
\begin{align}
	\mathrm\Upsilon_\text{av} \propto \langle v_\text{rel} \sigma \rangle \, N\ ,
\end{align}
where $\sigma$ is the relevant scattering cross-section, $v_\text{rel}$ the relative velocity of the two scattering particles,
and $N$ the number density of the scattering partner (for \SM particles this is just $N \sim T^3$). \footnote{%
Note that the feebly interacting particle does not have interact with the \SM at all, as the width $\mathrm\Upsilon_\text{av}$ can arise solely from interactions within the hidden sector.
} 

In the remainder of this section, we assume that $m \ll T_\star$ and focus on frequencies $f > f_\text{nr}$.
In this case, one has $m \ll T_\text{max}$, and we expect production at high temperatures ($m\ll T$) to dominate the overall \GWB.
Using the largely model-independent temperature scaling of $\mathrm\Upsilon_\text{av}$ at these temperatures, we can derive generic analytic approximations for the resulting \GWB.
However, we caution that, in principle, one also has to account for the non-relativistic contribution from temperatures $T < m$,which has to be evaluated on a case-by-case basis.

\subsubsection{Renormalizable interactions}
\label{sec:renormalisableinteractions}

If the width is dominated by renormalizable interactions with $d = 4$,
the physical momentum $k$ redshifts in the same way as the high-temperature width $\mathrm\Upsilon_\text{av}$.
The ratio $\nicefrac{\Uav}{k}$ is then fixed for a given frequency $f$, and the relativistic contribution to the background is always either in the Boltzmann (for $f \gg f_c$), hydrodynamic ($f\ll f_c$),
or intermediate regime (for $f \sim f_c$).
The production rate saturates the upper bound \eqref{eq:high temp max} at the critical frequency $f_c$,
which is the frequency at which the associated, redshifted graviton momentum satisfies condition \eqref{eq:maximal momentum},
\begin{align}\label{eq:critical frequency}
k&= \sqrt{10} \mathrm\Upsilon_\text{av} \quad \Leftrightarrow \quad
f = f_c \ , &
\frac{f_c}{y} &\equiv \frac{\sqrt{10}}{\pi} f_\text{peak}
\simeq \unit[6 \cdot 10^{10}]{Hz} \ .
\end{align}
The effective coupling $y$ is a combination of coupling constants and other model parameters.
Employing the same general assumptions as in section \ref{sec:present spectrum}, it is the only remaining model-dependent quantity,
and the critical frequency $f_c$ directly probes its size.
If one requires \eg $f_c \sim \unit[1]{Hz}$, so that the \GWB is maximal in the frequency window of the next-generation of laser interferometers,
this gives an optimal coupling parameter $y \sim 1.7 \cdot 10^{-11}$.

We estimate the high-temperature background by inserting the average width \eqref{eq:width scaling} into
the relativistic \GW production rate \eqref{eq:relat rate} in order to compute the integral in \eqref{eq:gravitational wave spectrum}.
For $f > f_\text{nr}$, this yields
\begin{align}\label{eq:renorm gw background}
h^2 \Omega_\text{gw}^{T>m} (f) &\overset{d=4}{\simeq}
1.6 \cdot 10^{-40} \times g_X \frac{f f_c}{f_c^2 + f^2}
\left( \frac{f}{\unit{Hz}} \right)^2
\min\left( \frac{f}{\unit{Hz}},\frac{f_\star}{\unit{Hz}} \right) \ .
\end{align}
As it should, this expression saturates the bound \eqref{eq:high temp max} for $f = f_c$.
$T_\star$ is the earliest temperature at which hidden sector thermal fluctuations begin to produce \GWs and, in principle, a free parameter.
As mentioned above, the hidden sector can thermalize well before or after the \SM,
so that $T_\star$ may be significantly lower than the maximal temperature of the \SM plasma after reheating.
However, the same processes that give rise to $\Uav \sim y T$ can also produce hidden particles via thermal freeze-in, which puts an effective lower bound on $T_\star$.
This production mode equilibrates when $\mathrm\Upsilon_\text{av} > H = \nicefrac{T^2}{m_0}$,
which gives $T_\star > y \, m_0$ (or equivalently $\sqrt{10} f_\star > f_c$) as the temperature where freeze-in production alone starts to be sufficient to thermalize the hidden sector.

Expression \eqref{eq:renorm gw background} is  generic and also captures the \SM contribution to the final \GWB.
Comparing the production rate \eqref{eq:sm gw production rate} with the width parametrization \eqref{eq:width scaling}, we find the critical frequency
\begin{align}\label{fcritSM}
f_c^\text{\SM} \approx \unit[3.1 \cdot 10^8]{Hz} \ .
\end{align}
This implies that \eqref{eq:sm gw production rate}, which is accurate in the hydrodynamic limit, is not expected to be valid for frequencies above $\unit[10^8]{Hz}$.
For smaller frequencies, and using $g_X = g_{\SM} = 6$, we find the \SM background
\begin{align}\label{eq:final sm background}
h^2 \Omega_\text{gw}^{\SM} (f) &\overset{f < f_c^{\SM}}{\simeq}
3.2 \cdot 10^{-48} \times \left( \frac{f}{\unit{Hz}} \right)^3 \min\left( \frac{f}{\unit{Hz}},\frac{f_\star}{\unit{Hz}} \right) \ .
\end{align}
While this background was already estimated in \cite{Ghiglieri:2015nfa,Ghiglieri:2020mhm,Ringwald:2020ist},
these references neglect that \GW emission is suppressed for super-horizon modes, delaying \GW production until after horizon re-entry.
We phenomenologically incorporate this effect by rescaling the maximal temperature according to \cref{eq:Tmax interp}.

\subsubsection{Non-renormalizable interactions}
\label{sec:nonrenorm interactions}

One way to enhance the production of the hidden particles at high temperatures is to include non-renormalizable interactions with $d>4$, which are more efficient at high temperature.
These interactions contribute to and potentially dominate the width in \cref{eq:width scaling}, resulting in more complicated temperature dependence compared to the renormalizable case.
In this case, the ratio $\nicefrac{\mathrm\Upsilon_\text{av}}{k}$ also depends on temperature,
so that a given \GW frequency $f$ can receive contributions from several different production regimes.

\begin{figure}
\includegraphics[width = 0.99\textwidth]{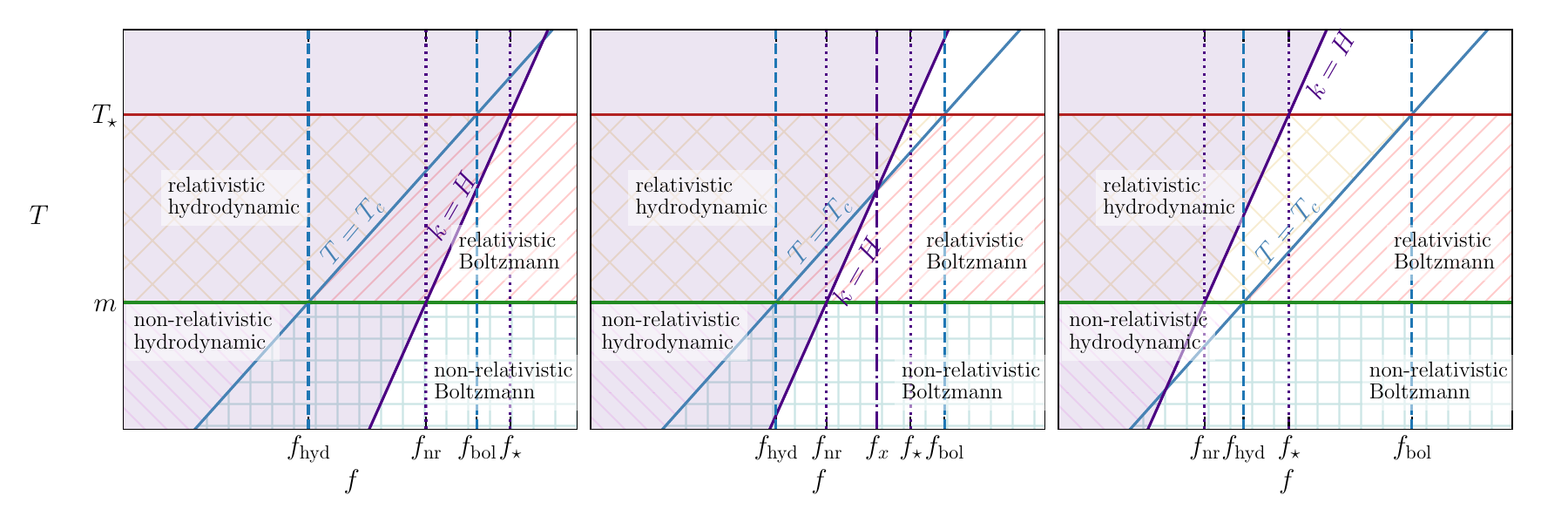}
\caption{\label{fig:GWregimes}
Cases of \GW production for a width $\Uav$ dominated by non-renormalizable interactions.
At each temperature, modes are produced either in the hydrodynamic ($k < \Uav$) or Boltzmann ($k > \Uav$) regime.
They transition from the hydrodynamic to the Boltzmann regime as the temperature drops below their respective critical temperature $T_c$.
Modes with $f>f_\text{bol}$ are only produced in the Boltzmann regime while modes with $f < f_\text{hyd}$ are only produced in the hydrodynamic regime.
The purple shading shows that \GW production is suppressed because modes are outside the horizon.
Modes with $f>f_\star$ are not affected by the horizon, while modes with $f<f_\mathrm{nr}$ enter it only after the hidden particle becomes non-relativistic.
For $T<m$, the transition between the hydrodynamic and Boltzmann regimes becomes more model-dependent, the line shown in the figure corresponds to $n=0$ from eq.~\eqref{eq:width scaling}.
}
\end{figure}

\Cref{fig:GWregimes} illustrates the interplay between these regimes and the horizon entry temperature, leading to a variety of \GW spectra.
As the universe expands and cools down, the width generically decreases faster than the momentum,
so that a \GW mode that is initially produced in the hydrodynamic regime can later transition to the Boltzmann regime.
For a given frequency $f$, and using the high-temperature scaling of the width $\mathrm\Upsilon_\text{av}$ in \eqref{eq:width scaling},
this transition occurs at the critical temperature
\begin{align} \label{eq:Tcriticaldefinition}
T_c(f) &= \Lambda \left(\frac{f}{f_c} \right)^{\nicefrac1{2(d-4)}} \ ,
\end{align}
where $f_c$ is the same critical frequency as in \cref{eq:critical frequency}.
The critical temperature increases with frequency and is equal to the cutoff scale $\Lambda$ for $f = f_c$.
Each other mass scale defines its own characteristic frequency as the frequency for which $T_c$ is equal to the mass scale.
In particular, the maximal temperature $T_\star$ corresponds to the Boltzmann frequency
\begin{align}\label{eq:fbol def}
f_\text{bol} &\equiv f_c \left( \frac{T_\star}{\Lambda} \right)^{2(d-4)} \ , &
T_c(f_\text{bol}) &= T_\star \ .
\end{align}
Above this frequency, $T_c$ is larger than $T_\star$, and the high-temperature background is produced entirely in the Boltzmann regime.
If the hidden particle is massive, its mass $m$ likewise corresponds to the hydrodynamic frequency
\begin{align}\label{eq:fhyd def}
f_\text{hyd} &\equiv f_c \left( \frac{m}{\Lambda} \right)^{2(d-4)} \ , &
T_c(f_\text{hyd}) &= m \ .
\end{align}
Below this frequency, $T_c$ is smaller than $m$, and the high-temperature background is produced entirely in the hydrodynamic regime.

Since we focus on the case $m \ll T_\star$, and since $m \ll \Lambda$ is necessary for a consistent \EFT description, $f_\text{hyd}$ has to be much smaller than both $f_\text{bol}$ and $f_c$.
Naively, one might assume that the \EFT setup also requires $T_\star \ll \Lambda$, and therefore $f_\text{bol} \ll f_c$, but this is not entirely true.
Considering larger maximal temperatures, the setup can be salvaged for modes with frequency $f < f_c$, where $T_c$ is smaller than $\Lambda$.
For temperatures $T_c < T < \Lambda$, these modes are first produced in the hydrodynamic regime.
Since $d > 4$, the resulting \GW production rate increases as the temperature decreases, so that the \GWB contribution at $T \sim T_c$,
where production transitions into the Boltzmann regime,
is more important than the contribution generated at temperatures $T \gtrsim \Lambda$, where the \EFT description breaks down.
Unfortunately, the same reasoning does not extend to frequencies $f > f_c$, which, at temperatures suitable for the \EFT description, are only produced in the Boltzmann regime.
Since the \GW production rate in the Boltzmann regime becomes smaller as the temperature decreases,
the final background is dominated by the contribution generated at the largest available temperature, \ie $T_\text{max}$.
Hence, and in line with the naive expectation for $T_\text{max} > \Lambda$, the \EFT description cannot be used to determine the size of the overall background at these frequencies.

\begin{figure}
\centering
\includegraphics[width=.49\textwidth]{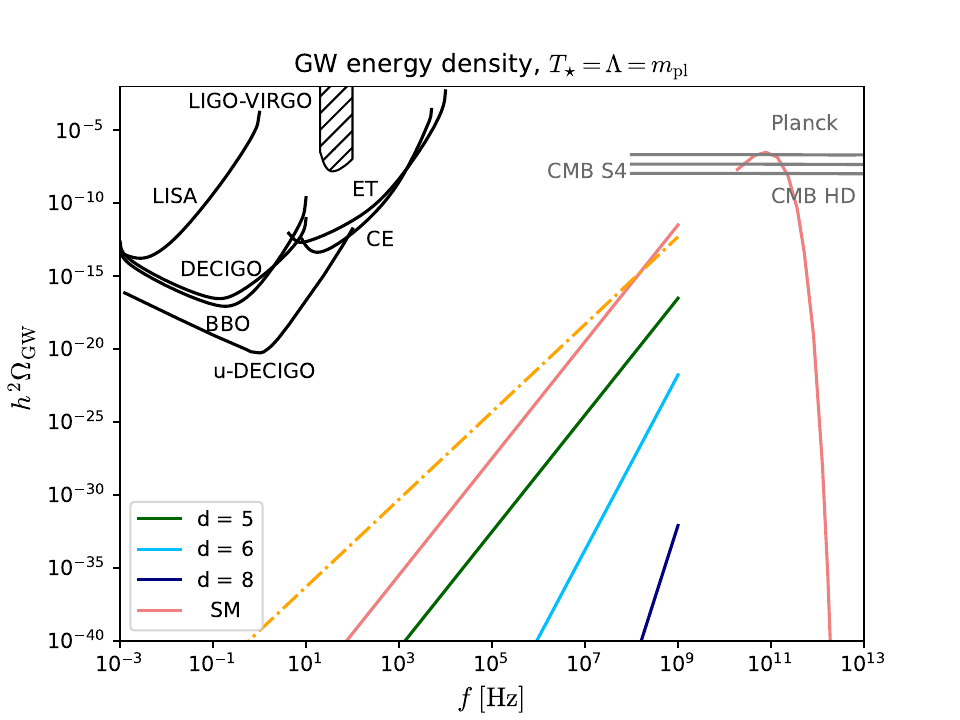}
\includegraphics[width=.49\textwidth]{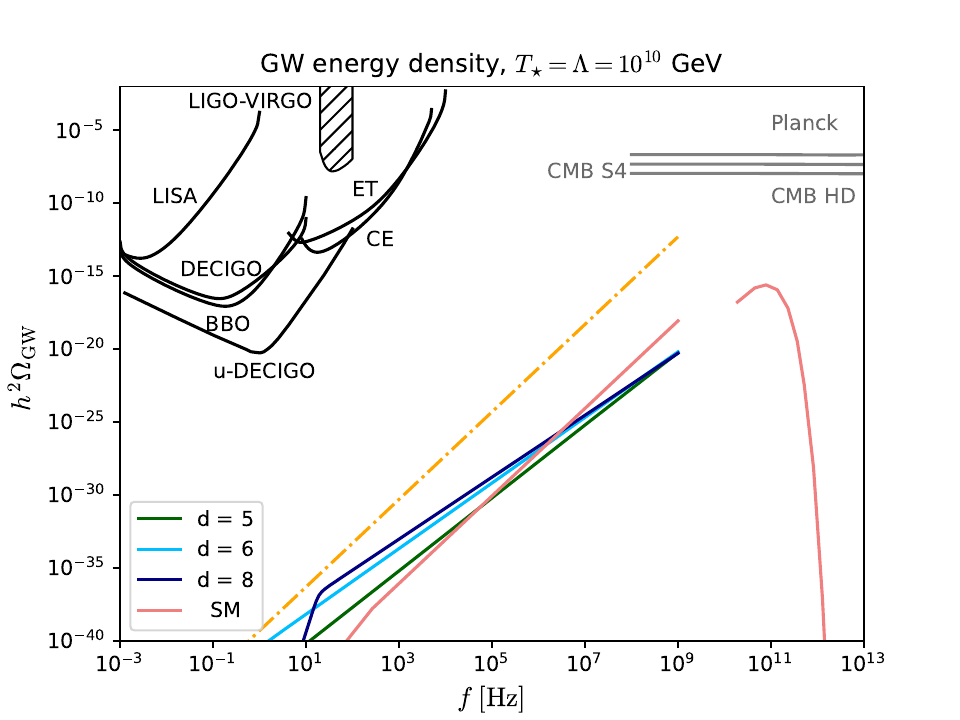}
\caption{\label{fig:GWdensity_NRinteractions}
Estimate of the \GW background from non-renormalizable interactions with dimension $d > 4$
for $T_{\star} = \Lambda = m_{\text{Pl}}$ (\textit{Left}) and $T_{\star} = \Lambda = 10^{10}$ GeV (\textit{Right}), \cf~\cref{eq:NRspectrumCase1,eq:NRspectrumCase2,eq:NRspectrumCase3}.
The dashed orange line shows the upper bound \eqref{eq:f small max} and the red line shows the \SM background resulting from the same value of $T_\star$.
The \SM line has been computed using the production rate from \cite{Ghiglieri:2015nfa} while accounting for previously missed super-horizon effects, see also the last paragraph in \cref{sec:renormalisableinteractions}.
Although it appears to violate the upper bound, which does in principle apply to the \SM contribution,
this occurs above the frequency \eqref{fcritSM}, where the validity of  \eqref{eq:final sm background} is questionable.
For much larger frequencies (beyond the visible gap) one enters the 
peak region, where vertex-type and other diagrams should be included.
This possibly also explains the apparent slight mismatch between the two parts of the curve.
In both plots, we consider a hidden particle mass $m=0$ and coupling parameter $y=1$, \cf~\cref{eq:width scaling}.
The signal can be enhanced and/or present additional features at low frequencies if $y\ll 1$, \cf~\eg~\cref{fig:ALPGWcontribution}. 
The smooth black lines are extracted from \cite{Schmitz:2020syl,Ringwald:2020vei} and show experimental power-law integrated sensitivities \cite{Thrane:2013oya} of various prospective laser interferometers. The black hatched region highlights the already excluded parameter space from the third observing run of LIGO and VIRGO \cite{KAGRA:2021kbb}.
Finally, the grey lines show the sensitivity of Planck \cite{Planck:2018vyg} and its proposed successors, CMB-S4 \cite{CMB-S4:2016ple} and CMB-HD \cite{CMB-HD:2022bsz}.
}
\end{figure}

At intermediate frequencies $f_\text{hyd} < f < f_\text{bol}$, 
the high-temperature background can be produced both in the hydrodynamic and Boltzmann regime,
based on the precise relationship between $T_c$ and $T_\text{entry}$.
Both characteristic temperatures are functions of $f$ and equal to each other at the crossing frequency
\begin{align}\label{eq:fcross def}
f_x &\equiv f_c  \left( \frac{f_\text{peak} \Lambda}{\pi \, f_c \, m_0} \right)^{1+\frac1{2(d-4)-1}} \ ,
& T_c(f_x) &= T_\text{entry}(f_x) \ .
\end{align}
The ordering of the characteristic frequencies $f_x$, $f_\text{hyd}$, and $f_\text{bol}$ determines the spectrum of the final \GWB.
Following the derivation outlined in \cref{sec:gwb appendix}, we find three physically distinct spectra:
\begin{enumerate}
\item 
If $f_\text{bol} < f_x$,  the horizon entry temperature is always smaller than the critical temperature.
Hence, \GW production is always in the Boltzmann regime, as shown in the left panel of~\cref{fig:GWregimes}.
The resulting spectrum has two knees, one at $f_\star$, and another one at $f_\text{nr}$. The high-temperature contribution to the \GWB is
\begin{align}
\label{eq:NRspectrumCase1}
h^2 \Omega_\text{gw}^{T > m}(f) &\overset{f_x > f_\text{bol}}{\simeq}
g_X \frac{1.6 \cdot 10^{-40}}{2(d-4) + 1} \left( \frac{f}{\unit{Hz}} \right)^2
\frac{f_\mathrm{bol}}{\unit{Hz}}
\begin{cases}
\quad 0 & \text{for} \quad f < f_\mathrm{nr} \, ,\\
\left(\frac{f}{f_\star} \right)^{2(d-4)} & \text{for} \quad f_\mathrm{nr} < f < f_\star \, , \\
\quad \frac{f_\star}{f} & \text{for} \quad f_\star< f \, .
\end{cases}
\end{align}
The background produced at lower temperatures $T<m$ contribution is highly suppressed and bounded by the general upper bound \eqref{eq: fixed mass bound}.
\item
If  $f_\text{hyd} < f_x < f_\text{bol}$, the high-temperature background receives contributions from both the Boltzmann and hydrodynamic regimes.
High frequencies $f>f_\text{bol}$ are produced in the Boltzmann regime and directly probe $T_\star$.
Intermediate frequencies $f_x < f < f_\text{bol}$ are produced in both regimes, since their critical temperature is smaller than the horizon entry temperature.
Since the critical temperature falls below the horizon entry temperature for lower frequencis $f < f_x$, these again only receive contributions from the Boltzmann regime.
Hence, the high-temperature contribution to the \GWB is
\begin{align}
\label{eq:NRspectrumCase2}
h^2 \Omega_\text{gw}^{T > m}(f) &\overset{f_\star \gg f_\text{bol}}{\simeq}
g_X \frac{1.6 \cdot 10^{-40}}{2(d-4) + 1} \left( \frac{f}{\unit{Hz}} \right)^2
\frac{f_\mathrm{bol}}{\unit{Hz}}
\begin{cases}
\quad 0 & \text{for} \quad f < f_\mathrm{nr}\,,\\
\left(\frac{f}{f_\star} \right)^{2(d-4)} & \text{for} \quad f_\mathrm{nr} < f < f_x\,, \\
\beta \frac{f_x}{f_\mathrm{bol}} \left(\frac{f}{f_x} \right)^{\nicefrac{1}{2(d-4)}} & \text{for} \quad f_x < f < f_\mathrm{bol}\,, \\
\quad \frac{f_\star}{f} & \text{for} \quad f_\mathrm{bol}< f\,.
\end{cases}
\end{align}
This expression coincides with \cref{eq:NRspectrumCase1}, except in the range $f_x < f < f_\mathrm{bol}$ that also receives a contribution from the hydrodynamic regime.
The numerical factor $1 \lesssim \beta \leq 4$ arises due to the integration over $T>T_c$.
\item
Finally, if $f_x < f_\text{hyd}$, the effects of the horizon can be almost completely neglected, leading to a \GWB with one knee at $f=f_\star$, one at $f=f_\mathrm{hyd}$, and a final one at $f_\mathrm{nr}$.
The high-temperature contribution to the \GWB is
\begin{align}
\label{eq:NRspectrumCase3}
h^2 \Omega_\text{gw}^{T > m}(f)
&\overset{f_\text{bol} \gg f_\star, f_\text{hyd} \gg f_\text{nr}}{\simeq}
g_X \frac{1.6 \cdot 10^{-40}}{2(d-4) + 1} \left( \frac{f}{\unit{Hz}} \right)^2
\frac{f_\mathrm{bol}}{\unit{Hz}}
\begin{cases}
\quad 0 & \text{for} \quad f < f_\mathrm{hyd}\,,\\
\beta \frac{f_x}{f_\mathrm{bol}} \left(\frac{f}{f_x} \right)^{\nicefrac{1}{2(d-4)}} & \text{for} \quad f_\mathrm{hyd} < f < f_\mathrm{bol}\,, \\
\frac{f_\star}{f} & \text{for} \quad f_\mathrm{bol}< f\,.
\end{cases}
\end{align}
\end{enumerate}

\Cref{fig:GWdensity_NRinteractions} shows the \GWB arising from a hidden particle whose width
is dominated by non-renormalizable interactions in comparison to the sensitivities of various prospective \GW detectors.
Focusing on the most optimistic scenario, we choose $\Lambda = T_\star$ and consider
the two choices for the reheating temperature $T_\star = m_\text{Pl}$ and $T_\star = 10^{10}$ GeV.
As benchmarks, we set $y=1$ and consider operators of dimensions $5$, $6$ and $8$.
While choosing smaller values for the coupling $y$ might enhance the signal, in general, virtual graviton exchanges constitute
an unavoidable contribution to the feebly interacting particles' width and effectively lead to a lower bound on the value of $y$.
Conservatively, \ie neglecting any quantum gravity effects at the Planck scale, we expect that this should contribute to the width
at the very least at dimension $8$, such that $y \sim \mathcal{O}(1)$ for $d\gtrsim 8$, which is the value we consider in the figure.

The plot illustrates the frequency dependence of the signal in line with the estimates in \cref{eq:NRspectrumCase1,eq:NRspectrumCase2,eq:NRspectrumCase3}.
For our choice of benchmarks, $f_\text{bol} \approx 6\cdot 10^{10}$ Hz lies above the range of validity of our approximation.
For the first benchmark with $T_\star = m_{\text{Pl}}$, one has $f_\star\approx \unit[3\cdot 10^{11}]{Hz}$.
Hence, $f<f_{\text{bol}}<f_x$ in our region of interest, and the \GW signal is captured by \cref{eq:NRspectrumCase1}. 
As expected, the \GWB in this scenario scales as $f^{2(d-3)}$ for all frequencies. The fact that $f<f_\star$ also explains why the \GWB is suppressed compared to our second scenario $T_\star = \unit[10^{10}]{GeV}$.
Indeed, while $T_\star=m_\text{Pl}$, the maximal temperature $T_\text{max}$ for which \GWs can be produced is much lower than the Planck scale and,
given that \GWs are always produced in the Boltzmann regime, \GW production is suppressed by $\left(\nicefrac{T_\text{entry}}{\Lambda}\right)^{2(d-4)}$.
The second benchmark $T_\star = 10^{10}$ GeV presents more interesting features.
In that case, one has $f_\star<f_\text{bol}<f_x$. For $d=8$, we observe that the crossing frequency $f_x$ lies in the LISA frequency band. Hence, the \GWB scales as $f^{2(d-3)}$ for $f\lesssim \unit[100]{Hz}$ and $f^{2+1/(2(d-4))}$ for larger frequencies,
as expected from eq.~\eqref{eq:NRspectrumCase2}. A similar feature is present for $d=5$ and $d=6$ but $f_x$ is pushed towards smaller frequencies.
Therefore, a measurement of the slope of the \GWB can be used, in principle, to deduce the mass-dimension of the interaction generating the width $\mathrm\Upsilon_\text{av}$.
Moreover, depending on the value of $T_\star$, it is in theory possible to measure the values of all the characteristic frequencies $f_\text{nr}$, $f_\text{hyd}$, $f_x$, $f_\text{bol}$ and $f_\star$ by looking at variations in the slope of the \GWB.
This way, one may be able to extract information on the mass of the particle as well as the the \UV scale $\Lambda$ and reheating temperature $T_\star$.

\section{Explicit examples}\label{Sec:Examples}

In the following we illustrate our results for two popular extensions of the SM,
namely axion-like particles (ALP) and heavy neutral leptons (HNL). 
 
\subsection{Axion-like particles}

First appearing as a solution to the strong CP-problem \cite{Peccei:1977ur,Peccei:1977hh,Wilczek:1977pj,Weinberg:1977ma} (for a review see \eg \cite{DiLuzio:2020wdo})
\footnote{It has recently been pointed out that the phases from instanton effects and from the mass term are systematically aligned if one reverses the order of the sum over topological sectors and the infinite volume limit,\ie,
if these limits are taken the other way around, there is no physical phase, suggesting that there is actually no strong CP problem \cite{Ai:2020ptm}.} 
axions and, more generically, \ALPs, are also viable 
dark matter candidates \cite{Preskill:1982cy,Dine:1982ah,Abbott:1982af} (for a review see \eg \cite{Adams:2022pbo}),
and can drive inflation \cite{Freese:1990rb}, including realistic warm inflation scenarios \cite{Berghaus:2019whh,Laine:2021ego}. 
As a result, \ALPs have become one of the most popular extension of the \SM  and are currently searched for at a wide range of different
experiments \cite{Graham:2015ouw,Sikivie:2020zpn,Agrawal:2021dbo,Adams:2022pbo,Antel:2023hkf,Blondel:2022qqo,Antel:2023hkf} and astrophysical environments  \cite{Raffelt:2006cw,Agrawal:2021dbo}.

\begin{figure}
\centering
\includegraphics[width=.49\textwidth]{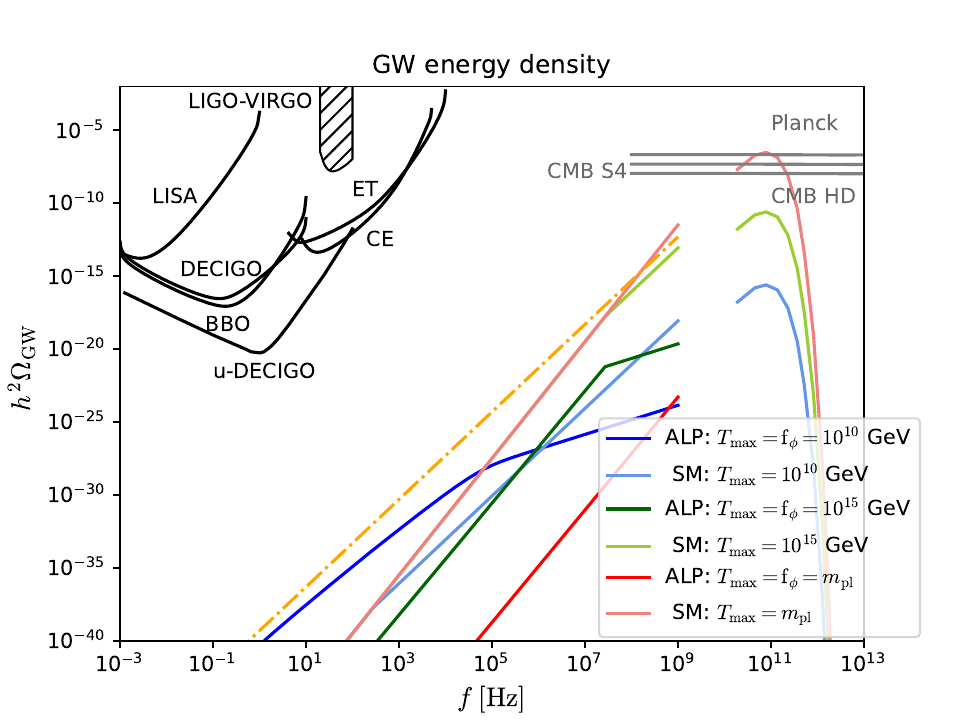}
\includegraphics[width=.49\textwidth]{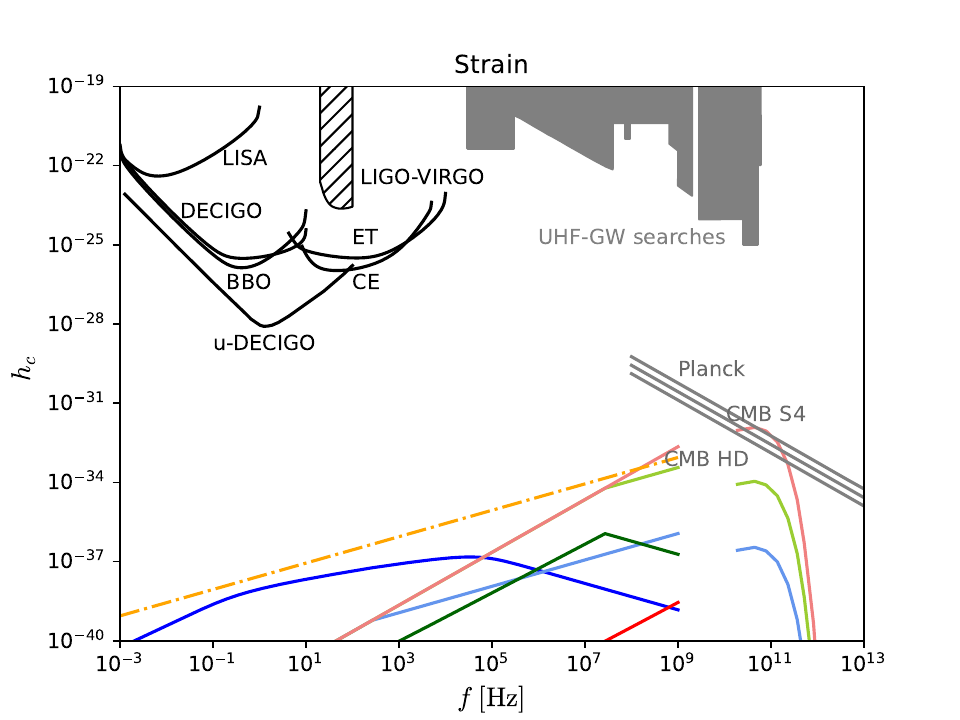}
\caption{\label{fig:ALPGWcontribution} \GWB due to thermal fluctuations of a massless \ALP with axion decay constants $\mathrm f_\phi \in \{ 10^{10} \mathrm{GeV},10^{15} \mathrm{GeV}, m_{\text{Pl}}\}$,
as measured by the energy density $h^2 \Omega_\text{gw}$ (\textit{Left}) or strain $h_c$ (\textit{Right}).
For each line, we choose $T_\star = \mathrm f_\phi$
and $\Lambda_{\text{IR}}\approx \unit[200]{MeV}$, considering QCD-like gauge field.
The corresponding \SM background for the same reheating temperature $T_\star$ is shown in a lighter version of the same color (blue, green and red respectively).
The orange dashed line shows the upper bound \eqref{eq:f small max} (for $g_X = 1$).
For details about the \SM line and the various experimental constraints in the LISA frequency band, see the caption below \cref{fig:GWdensity_NRinteractions}. Additionally, we display in the right plot the sensitivity estimate of some recently proposed searches\protect\footnotemark~for high-frequency \GWs \cite{Arvanitaki:2012cn,Goryachev:2014yra,Aggarwal:2020umq,Aggarwal:2020olq,Domcke:2020yzq,Ringwald:2020ist,Berlin:2021txa,Domcke:2022rgu} (\cf also \cite{Berlin:2023grv}).
}
\end{figure}
\footnotetext{We caution that these searches rely on different sets of assumptions, \textit{e.g.} about the strength of the cosmic magnetic fields, and show different levels of maturity.
Hence, the gray region might have to be slightly adapted in the future.}
While the \GWB from \ALPs was considered within the SMASH \cite{Ballesteros:2016xej} model in \cite{Ringwald:2020ist},
only the impact on the reheating temperature\footnote{%
Some of the most studied axion production mechanisms, \cf~\eg the misalignment mechanism \cite{Preskill:1982cy,Abbott:1982af,Dine:1982ah},
are non-thermal, such that $T_{\star}$ effectively corresponds more to a thermalization temperature than the reheating temperature.}
$T_\star$ was accounted for; the enhancement due to the feebly-interacting nature of ALPs as well as the suppression coming from the requirement for modes to remain sub-horizon were missed. 

In gauge theories, \ALPs are generically expected to couple to the topological charge density of the gauge sector,
\begin{align}
	\label{eq:alp coupling}
	\mathcal L &\supset \frac12 \left( \partial_\mu \phi \partial^\mu \phi - m^2 \phi^2 \right) - \frac{\phi\,\chi}{\mathrm f_\phi} \ , &
	\chi &= \frac{\alpha}{16\pi} \widetilde G_{\mu\nu} G^{\mu\nu} \ ,
\end{align}
where $\phi$ is the axion-like particle, $G_{\mu\nu}$ the gauge field strength tensor, and $\mathrm f_\phi$ the axion decay constant.
In a minimal setup, $G_{\mu\nu}$ can be \eg identified with the \SM gluon field strength tensor,
but one may also consider theories with hidden gauge fields that are not as strongly constrained by \SM observations.
In the following, we primarily focus on the case of very light \ALPs, with mass $m \ll T$.
In this case, non-perturbative effects are important for an accurate estimate of the width $\mathrm\Upsilon_\text{av}$.
For an asymptotically free theory, one obtains \cite{Laine:2022ytc}
\begin{align}\label{eq:alp width}
	\mathrm\Upsilon_\text{av} &\overset{m\ll T}{=} \kappa \, n_c^3 (n_c^2 - 1) \frac{\alpha^5 T^3}{\mathrm f_\phi^2} \ , &
	\kappa &\approx 1.5 \ , &
	\frac1{\alpha} &\approx \frac{22 n_c}{12\pi} \ln \left( \frac{2\pi T}{\Lambda_\text{IR}} \right) \ ,
\end{align}
where $n_c$ is the number of colours associated with the gauge fields
and $\Lambda_\text{IR}$ the infrared confinement scale at which the gauge coupling $\alpha$ exhibits a Landau pole.
In contrast to the generic discussion in \cref{Sec:AnalyticEstimates}, we here also account for the running of $\alpha$.
For $n_c = 3$, one finds
\begin{align}
f_\text{bol} &\approx \left(\frac{T_\star}{\mathrm{f}_\phi}\right)^2\left( \frac{\alpha(T_\star)}{0.02} \right)^5 \times \unit[6 \cdot 10^{4}]{Hz}.
\end{align}
$f_x$ can be obtained numerically but does not have an analytical expression.
Indeed, the logarithmic temperature dependence of the coupling $\alpha$ is not captured by eq.~\eqref{eq:width scaling} rendering the approximation \eqref{eq:fcross def} inadequate.

Figure \ref{fig:ALPGWcontribution} shows the \GWB predicted by the approximate formula \eqref{eq:relat rate} for various values of $\mathrm f_\phi$.
We show the results for both the \GW energy density fraction $\Omega_{\text{gw}}$ per logarithmic wave number interval  and the characteristic strain \eqref{eq:gw strain}.
The reheating temperature $T_\star$ is, in principle, an independent parameter.
We consider the case $T_\star = \mathrm{f}_\phi$, 
which is the maximal reheating temperature for which our EFT approach remains valid over the entire frequency range.
Smaller reheating temperatures $T_\star\ll \mathrm{f}_\phi$ result in a suppressed \GWB.

The figure reproduces the behaviour expected from the approximations \eqref{eq:NRspectrumCase1,eq:NRspectrumCase2,eq:NRspectrumCase3}, up to small modifications due to the running of $\alpha$.
For the first two benchmarks $T_\star=m_{\text{Pl}}$ and $T_\star = \unit[10^{15}]{GeV}$, one has $f_x > f_\text{bol}$. In this case, the spectrum is determined solely by the position of the horizon frequency $f_\star$, see eq.~\eqref{eq:NRspectrumCase1}.
For $T_\star=m_{\text{Pl}}$, $f_\star$ lies above the range of validity of our approximations, so that $\Omega_\text{gw}$ approximately scales as $f^4$.
On the other hand, for $T_\star = \unit[10^{15}]{GeV}$, $f_\star \approx \unit[3\cdot 10^7]{Hz}$ lies well within our region of interest.
In this case, the \GWB scales first as $f^4$ for $f<f_\star$, and then as $f$ for $f>f_\star$. The third benchmark $T_\star = 10^{10}$ GeV is slightly more complex since $f_x<f_{\text{bol}}$, and is captured by eq.~\eqref{eq:NRspectrumCase2}.
Hence, the spectrum exhibits two kinks at $f=f_x\approx \unit[0.1]{Hz}$ and $f=f_{\text{bol}}\approx \unit[9\cdot 10^4]{Hz}$.
Below $f_x$, the signal grows as $f^4$, between $f_x$ and $f_\text{bol}$, it grows as $f^{5/2}$, and above $f_\text{bol}$, it grows linearly. 

Additionally, we observe that only the case $T_\star = \unit[10^{10}]{GeV}$ can overcome the equivalent \SM background with the same value of $T_\star$.
Indeed, for $T_\star = \unit[10^{15}]{GeV}$ and $T_\star = m_{\text{Pl}}$, we always have $T_\text{entry}<T_c$ and, hence, \GWs are always produced in the Boltzmann regime.
For our third benchmark $T_\star = \unit[10^{10}]{GeV}$, hydrodynamic contributions dominate the \GW production for $f<0.1$ Hz and the \ALP \GWB can exceed the \SM contribution by $\sim 10$ orders of magnitude in the LISA frequency band.
However, even though it can surpass the \SM signal (for the same value of $T_{\mathrm{max}}$) by several orders of magnitude,
the \ALP contribution remains much too small for any hope of detection at proposed laser interferometer or high frequency \GW experiments.
We also observe that constraints from $N_{\mathrm{eff}}$ are much too weak in that frequency band to constrain hidden sectors.
The latter constraints become relevant in the Boltzmann regime  \cite{Ghiglieri:2020mhm,Ringwald:2020ist},
where the \ALP contribution is subdominant compared to the \SM background \cite{Klose:2022knn}.

\subsection{Heavy neutral leptons}
\label{sec:HNL}

Heavy neutral leptons (HNL) are a well-motivated extension of the \SM that can potentially explain several of its shortcomings \cite{Drewes:2013gca,Abdullahi:2022jlv},
including cosmological puzzles, such as the origin of the observed matter-antimatter asymmetry \cite{Canetti:2012zc} through leptogenesis \cite{Fukugita:1986hr,Akhmedov:1998qx,Asaka:2005pn}
or the dark matter puzzle \cite{Dodelson:1993je,Shi:1998km}, \cf~\cite{Garbrecht:2018mrp,Klaric:2021cpi} and \cite{Drewes:2016upu,Boyarsky:2018tvu} for summary articles.
An HNL with mass well below the GUT scale may interact only very feebly with the \SM.

\begin{figure}
\centering
\includegraphics[width=.49\textwidth]{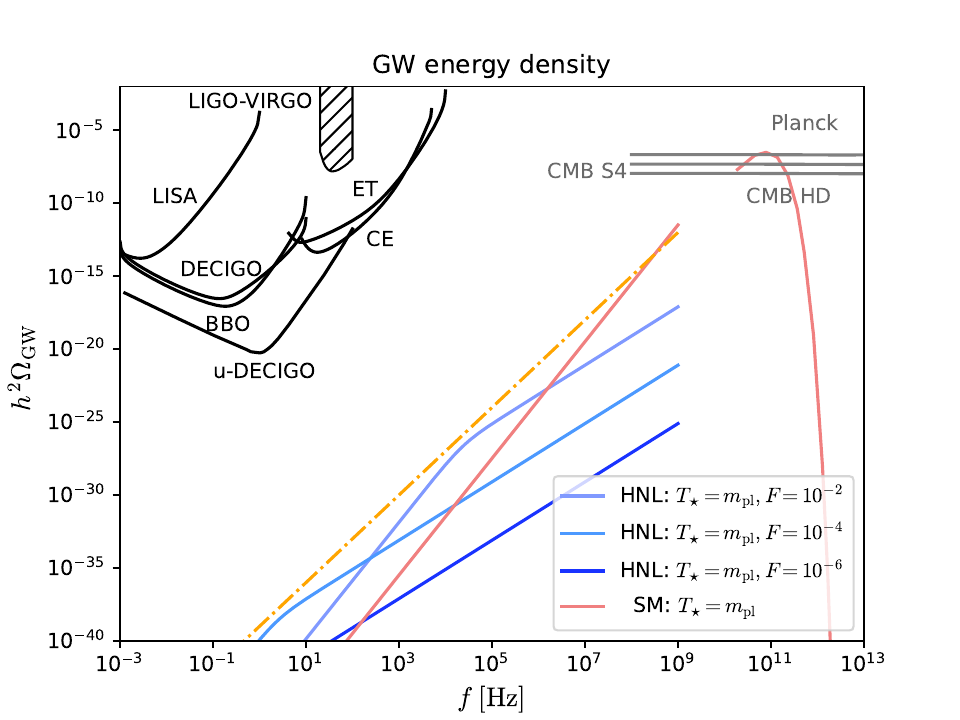}
\includegraphics[width=.49\textwidth]{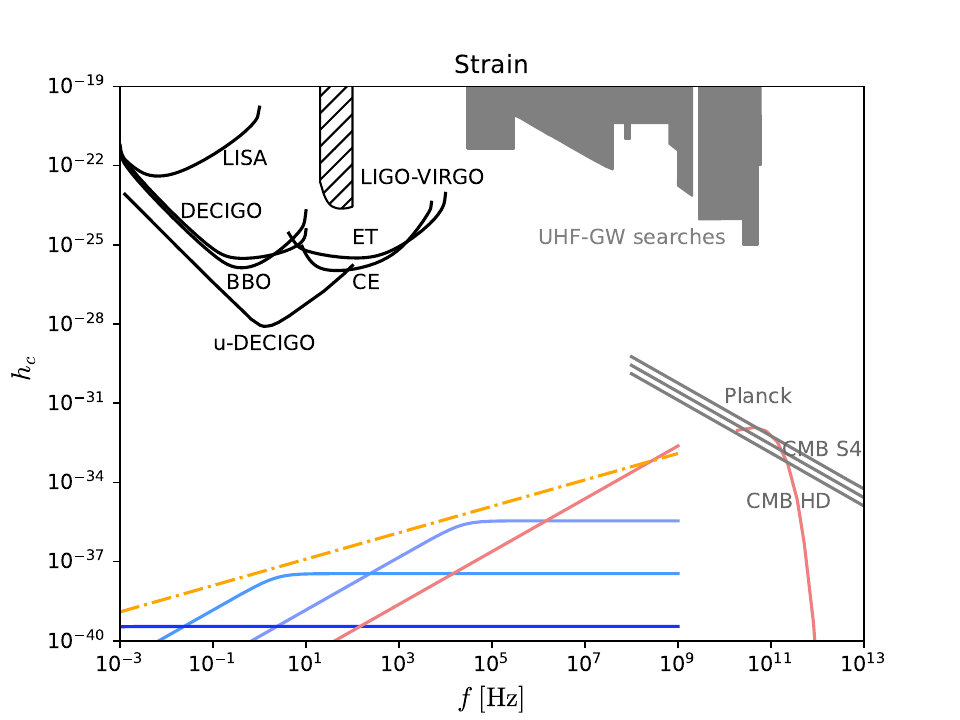}
\caption{
\GWB
due to thermal fluctuations of a relativistic \HNL in the early universe plasma,
as measured by the energy density $h^2 \Omega_{\rm{GW}}$ (\textit{Left}) and strain $h_c$ (\textit{Right}).
For details about the various experimental constraints, see the captions of \cref{fig:GWdensity_NRinteractions,fig:ALPGWcontribution}.
The light red line shows the expected \SM background for $T_{\mathrm{max}} = m_{\text{Pl}}$, see also the caption below \cref{fig:GWdensity_NRinteractions} for more details.
The remaining continuous coloured lines show the \HNL contribution for various parameter choices. 
Finally, the dashed orange line corresponds to the upper bound \eqref{eq:f small max}}
\label{fig:GWemissioncombined}
\end{figure}

Its contribution to the cosmological \GWB was estimated within the $\nu$MSM \cite{Asaka:2005pn} in \cite{Ringwald:2020ist} but,
similarly to the axion case, the impact of the Hubble horizon and the enhancement of its signal due its feeble interactions strength were not accounted for.
In the minimal type-I seesaw model, the only relevant interaction is
\begin{align}
	\mathcal L \supset F \psi (\tilde \phi^\dagger \ell) + \text{h.c.} \ ,
\end{align}
where $F$ is the \HNL Yukawa coupling, $\phi$ the \SM Higgs doublet, and $\ell$ a left-handed \SM fermion doublet.
The optical theorem connects the \HNL width $\mathrm\Upsilon_p$ to its production rate in the early universe plasma,
which has been studied extensively, see \eg 
\cite{Biondini:2017rpb}.
One finds that the momentum-averaged width is given as \cite{Garbrecht:2019zaa}
\begin{align}
\mathrm\Upsilon_\text{av} &= \gamma(z) \frac{F^2 T}{16\pi} \ ,
\end{align}
where $\gamma$ is a dimensionless function of $z \equiv \nicefrac{m}{T}$.
The precise determination of $\gamma$ is an involved computation, but for our purposes it is sufficient to use the simple approximation
\begin{align}
\gamma &\approx \max\left( 0.2, z-\frac32 \right) \ .
\end{align}
Evaluating the width in the high-temperature limit, this yields the critical frequency
\begin{align}
f_c &= \unit[2.4 \cdot 10^{8}]{Hz} \times F^2 \ .
\end{align}

Figure \ref{fig:GWemissioncombined} shows the \GW spectrum for different Yukawa couplings and an optimistic maximal temperature $T_\star = m_{\text{Pl}}$,%
\footnote{While the equilibration temperature in case of thermal production $T_\star \approx 4\cdot 10^{-3} m_0 F^2$ tends to be much lower than the Planck scale,
HNLs can also be produced at much higher temperature from \eg inflaton decay \cite{Giudice:1999fb,Shaposhnikov:2006xi}.}
comparing it with excepted sensitivities of various prospective gravitational wave detectors.
The hydrodynamic regime ($f < f_c$), where the \HNL \GWB exceeds the \SM contribution by several orders of magnitude, is clearly distinct from the 
the Boltzmann regime ($f > f_c$).
Although we observe a strong enhancement compared to the \SM background, which can be exceeded by up to 10 orders of magnitude,
the model independent upper bound \eqref{eq:f small max} forces the signal to lie much below any (currently proposed) detector sensitivities. 

\section{Discussion and conclusion}
\label{sec:conclusion}

In this work, we investigated \GW production from thermal fluctuations of hidden particles in the early universe. In contrast to more model-dependent \GWBs from \eg strong first order phase transitions,
reheating, or cosmic strings, this background does not rely on deviations from standard cosmology and is therefore much harder to avoid.

To estimate the size of the background, we used the closed time-path (Schwinger-Keldysh) 
formalism to compute the \GW production rate due to fluctuations of a feebly interacting fermion,
and combined this rate with prior results for (pseudo-)scalar particles in section \ref{sec:prodrates}.
Given the striking similarity of \cref{eq:gw_scalar_prodrate,eq:fermionic prodrate}, we expect that, up to an additional spin multiplicity factor,
\GW production due to fluctuations of a vector particle is
captured by an expression analogous to \cref{eq:gw_scalar_prodrate}.
We note in passing that the evaluation of the fermionic production rate at zero energy gives the corresponding contribution to the shear viscosity of the early universe plasma,
a quantity that is of interest well beyond its application to \GW production.

Using these rates, we then derived phenomenological formulae for \GWBs due to thermal fluctuations of generic feebly interacting particles
and obtained a model-independent upper bound on such backgrounds in \cref{sec:pheno}.
Our formulae incorporate the well-known fact that super-horizon modes of tensor perturbations are static.
This effectively delays the onset of \GW production until after horizon re-entry,
modifying some prior results for the \GWB predicted by the \SM \cite{Ghiglieri:2015nfa,Ghiglieri:2020mhm,Ringwald:2020ist}
(\cf \cref{eq:final sm background} and the discussion thereafter).
We also investigated backgrounds generated by particles whose width is dominated by non-renormalizeable interactions.
In principle, these backgrounds encode a plethora of information including  the mass of the emitting particle,
the dimension of the operator dominating its width, the corresponding \UV cutoff scale, and the maximal temperature of the hidden sector plasma.
Unfortunately, it seems impossible to observe such \GWBs from thermal fluctuations at LISA and other laser interferometers,
as the strain induced by the upper bound is still several orders of magnitude below the sensitivity of even beyond next generation experiments such as u-DECIGO.
The same applies to present efforts of detecting \GWs through electromagnetic conversion \cite{Domcke:2023qle} (\cf also \cite{Giovannini:2023itq,Giovannini:2023ceg}),
but this is still a young and very active field of research, and it is widely believed that the sensitivity can be increased by several orders of magnitude.\footnote{%
However, observing even \GWBs that saturate our upper bound has only been shown possible assuming that the interference term between signal and
background field in a cavity can be measured at the same accuracy as the signal itself \cite{Herman:2022fau}, and no method or technology to do this is known (\cf \cite{Navarro:2023eii} for a discussion).
}

In \cref{Sec:Examples}, we finally computed the thermal backgrounds from \ALPs and \HNLs  and discussed their properties.
We found that, in principle, measuring the different features of such backgrounds would give us access to various model parameters,
including the mass of the feebly interacting particle, its coupling strength, and the dimension of the operator dominating its width.
We also noticed that the resulting \GWBs can be enhanced by 5-10 orders of magnitude compared to the \SM background for $f\sim \mathcal{O}(1)$ Hz.
Despite this enhancement, these signals unfortunately remain orders of magnitude below the best sensitivity from next-generation laser interferometers or high-frequency \GW experiments,
consistent with the upper bound found in section \ref{Sec:UpperBound}.

Given the pessimistic implications of the upper bound, it is instructive to summarise and question the assumptions under which it was derived, and to assess how it may be avoided.
As a general comment, we in practice consider each new particle individually, meaning that our bound constrains the \GWB contribution per particle.
Its derivation relies on the following assumptions:
\begin{enumerate}
\item
\emph{Standard Cosmic History.} Our results apply for a standard cosmic history.
This amounts to two related but not identical assumptions.
\emph{(i)} In \eqref{eq:entropy conservation} and following we assume that \SM degrees of freedom dominate the effective number of radiation degrees of freedom.  
Relaxing this assumption does not  restrict the applicability of the method used here.
It would simply require tracking the entropy and temperature evolution in both sectors separately.\footnote{%
Note that increasing the number of degrees of freedom enhances the \GW production rate,
but this enhancement is overcompensated by the subsequent dilution.
As a result, models with more degrees of freedom (such as Grand Unified Theories, GUTs) tend to produce
a weaker signal than the \SM unless the contribution from new particles is enhanced by some mechanism,
such as the weakness of their interaction in the present work, \cf~section~\ref{Sec:Enhancement}.
However, GUTs can produce \GWBs in various other ways, \cf \cite{King:2021gmj,Fu:2023mdu}.}
\emph{(ii)} We assume that no entropy is transferred between the two sectors.
If entropy is transferred from the hidden sector to the \SM, the \GWB is diluted.
If entropy is transferred from the \SM to the hidden sector, the \GWB can be enhanced, but the details depend on the thermal history of the hidden sector.
We sketch how the formulae in section \ref{sec:gravitational waves from hydro} are generalised if these assumptions are relaxed in appendix \ref{sec:non standard temperature evolution appendix}.
\item
\emph{Focus on the hidden-sector-dominated contribution.}
We assume that our computation captures the dominant part of the \GWB from hidden sectors. 
This amounts to two separate assumptions, one of which is conceptual and the other technical in nature.
\emph{(i)} We consider one hidden particle at a time.
This is based on the observation that the \GWB at low frequencies tends to be dominated by the most weakly interacting particle \cite{Ghiglieri:2015nfa,Ghiglieri:2020mhm,Ringwald:2020ist}.
If there are several feebly coupled particles, 
the upper bound in section \ref{Sec:UpperBound} applies to each of them individually.
\emph{(ii)} In our computation of the \GW production rate due to a given hidden particle,
we focused on contributions from diagrams with a topology as in figure \ref{fig:wave_diag} (wave-function type diagrams)
with resummed propagators in the loop, which are subject to the enhancement discussed in section \ref{Sec:Enhancement}.
This enhancement is the largest for the particle with the weakest coupling, hence one expects this contribution to dominate in hidden sectors. 
Our computation
does not capture vertex-type corrections such as the diagram in figure \ref{fig:vertex_diag}. 
Since the enhancement is absent in the Boltzmann regime $k \gg \mathrm\Upsilon_\text{av}$, one can expect contributions from
vertex-type diagrams to be of the same order of magnitude as the ones considered here.
Hence, we expect order one corrections to \GW emission in the Boltzmann regime,
the precise computation of which in general may require further resummations (\cf \eg \cite{Anisimov:2010gy}). 
\item
\label{EqAssumpt}\emph{Equilibrium between the sectors.}
Throughout sections \ref{sec:pheno} and \ref{Sec:Examples} (and in particular in figures \ref{fig:GWdensity_NRinteractions}-\ref{fig:GWemissioncombined})
we assumed that the hidden sector is in equilibrium with the \SM at a temperature $T$. 
When this is not the case, one can distinguish two possibilities.
If the hidden sector is in a thermodynamic state that can be described by an effective temperature $T_h\neq T$,
then all references to the temperature in the distribution functions
have to be replaced with the corresponding values for $T_h$.
For example, this can be the case if hidden sector interactions are strong enough to maintain local thermal equilibrium within each sector,
but the portal interaction with the \SM is too weak to equilibrate the two sectors with each other.
When the hidden sector is in a true non-equilibrium state, 
our computation can still be applied when using \eqref{GeneralPi} instead of \eqref{eq:fermionic prodrate},
practically replacing the equilibrium distribution functions in the propagators with general phase space distribution functions,
provided that the loop integral computed in appendix \ref{Sec:FermionDiagram} is dominated by contributions from the dressed single (quasi)particle poles.
While this assumption can in principle be violated (e.g.~when the phase space distribution functions exhibit strong peaks,
collective excitations play a role, or the continuum contribution from branch-cuts dominates), it can be expected to hold for a wide range of non-equilibrium situations.     
An immediate consequence of this is that one expects an upper bound such as \eqref{eq:simplified max} to hold in non-equilibrium situations,
provided that the temperature is replaced by an effective parameter that characterises the occupation numbers.
Quantitatively this generalised upper bound can be weaker than the result \eqref{eq:simplified max} obtained in equilibrium.
\end{enumerate}
A detailed study on the impact of relaxing one or several of these assumption  quantitatively goes beyond the scope of this work.
We shall nevertheless briefly comment on what one may expect in two phenomenologically important situations.

\emph{Thermal freeze-in} describes the incomplete equilibration of some particle species due to a weak coupling and can be relevant for both \DM production
\cite{Dodelson:1993je,Hall:2009bx} and baryogenesis \cite{Akhmedov:1998qx,Asaka:2005pn,Klaric:2021cpi}. 
Freeze-in scenarios can involve a huge deviation from thermal equilibrium. However, since this is an \emph{under}abundance, the \GWB from the concerned particle is expected to be suppressed relative to our upper bound. 
Recall that the upper bound \eqref{eq:simplified max} is the result of a trade-off between two effects that apply irrespective of whether the hidden particle is in equilibrium or not: 
A smaller hidden particle width $\mathrm\Upsilon_p$ leads to a stronger enhancement of the \GWB per mode, but reduces the range of modes that are enhanced.  
During freeze-in (when the hidden particles are produced thermally) there is a third effect: A lower $\mathrm\Upsilon_p$ also implies a slower approach to equilibrium, hence lower occupation numbers. 
If those can be characterised by an effective temperature $T_h < T$, then it is straightforward to quantify the resulting suppression of the \GWB by replacing $T\to T_h$ in \eqref{eq:gw_scalar_prodrate} and \eqref{eq:fermionic prodrate};
if not, one has to use the equations in appendix \ref{Sec:FermionDiagram} with the full nonequilibrium distribution in \eqref{KadanoffBaym} and following.
At a qualitative level it is clear that an underabundance will suppress the contribution to the \GWB.

An overabundance, on the other hand, can potentially enhance the \GWB. An extreme case of this kind can be realised at the end of cosmic inflation. Indeed,
the results obtained in a number of recent works  studying graviton emission during reheating \cite{Kanemura:2023pnv,Barman:2023rpg,Barman:2023ymn,Bernal:2023wus} 
appear to violate the upper bounds formulated in section \ref{Sec:UpperBound}.
While the overabundance of inflaton occupation numbers implies that the upper bound should be relaxed,
these works appear to compute the \GW production rate without accounting for (i) the resummation of wave-function type corrections
that is necessary in the hydrodynamic regime (\cf~appendix~\ref{Sec:FermionDiagram}), and (ii) the suppression of \GW emission
for super-horizon modes that effectively delays the onset of \GW production until they re-enter the horizon.%
\footnote{This was previously pointed out in \cite{Tokareva:2883492,Tokareva:2023mrt}.}
We expect that systematically incorporating these effects would yield a significant suppression of the resulting \GWB.
However, it should also be added that some of the assumptions mentioned in point \ref{EqAssumpt} can be violated during reheating. In particular, inflaton decays can produce peaked distribution functions.
Moreover, the oscillating inflaton condensate generates effective masses for all particles that couple to it,
which can not only lead to non-perturbative particle production due to a parametric resonance, but also modify the perturbative production rate in a way that is not captured by the formulae presented here.
Finally, at the onset of reheating the density of particles is low, implying a long mean free path and hence a smaller $\mathrm\Upsilon_\text{av}$ than we used.
Hence, while the \GWB produced from reheating may exceed the upper bound presented here, its quantitative computation requires taking full account of all aforementioned effects and remains an open question.

In summary, we found that the thermal \GWB emitted by hidden sectors can dominate over the \SM contribution by orders of magnitude in the low frequency regime.
In spite of this, the upper bound from section \ref{Sec:UpperBound} and its potential non-equilibrium generalisations
imply that the background remains orders of magnitude below the sensitivity of any proposed interferometer
if the concerned particles are in thermal equilibrium or underabundant.
This pessimistic conclusion may be avoided in situations involving overabundances, such as cosmic reheating, but an accurate formalism to quantify this is still lacking.
Hence,  while the thermal \GWB for a given extension of the \SM in principle represents an unavoidable floor for \GW measurements,
this floor is so low that probing hidden sectors with \GWB measurements will almost certainly rely on more model-dependent sources in the foreseeable future. 

\section*{Acknowledgments}

The authors thank Anna Tokareva for making us aware that we forgot to implement the suppression on super-Hubble scales in the initial version of our plots
and Diego Blas for a discussion on \GW detection in resonant cavities.
Y.G. acknowledges the support of the French Community of Belgium through the FRIA grant No.~1.E.063.22F.

\appendix
\appendixpage
\addappheadtotoc

\section{GW production rate from fermions}\label{Sec:FermionDiagram}

To compute the fermionic equivalent of rate \eqref{eq:gw_scalar_prodrate},
we start from the Kubo formula \eqref{eq:kubo formula} and adapt the strategy used in \cite{Jackson:2018maa,Klose:2022knn}.
 The relevant contribution to the stress-energy tensor is
\begin{align}
T_{ij}(x) &\supset \frac{c_X}4 \overline \psi_x \i D_{ij} \psi_x \ , &
\i D_{ij} &= \gamma_i \i \partial_j + \gamma_j \i \partial_i \ ,
\end{align}
where $\psi_x = \psi(x) $ is either a Dirac or Majorana spinor, evaluated at $x = (t, \bm x)$, and the constant $c_X$ is $c_D = 2$ for a Dirac spinor and $c_M = 1$ for a Majorana spinor.
In the interest of full generality, we consider a setup in which the fermion is not necessarily in thermal equilibrium,
and use the Schwinger-Keldysh (In-In) formalism of non-equilibrium quantum field theory to evaluate the correlator in \eqref{eq:stress energy correlator}.
For reviews of this formalism, see \eg \cite{Cornwall:1974vz,Prokopec:2003pj,Berges:2015kfa,Garbrecht:2018mrp}.
The central idea is to evaluate the path-integral along a closed time-path (CTP) running from an initial time $t_0$ to a final time $t_f = + \infty$ and then back to $t_0$.
If the state of the system at $t=t_0$ is specified, the CTP path integral can then be used to compute path-ordered, out-of-equilibrium correlation functions
\begin{align}
\langle \T_C \left\{ \mathcal O_1^{a_1} \dots \mathcal O_n^{a_n} \right\} \rangle \ ,
\end{align}
where $\T_C$ is the time-path ordering operator and the $\mathcal O_i^a(x_i)$ are local operators that depend only on field operators evaluated at $x_i$.
Each field operator, and with it each composite operator $\mathcal O_i^{a_i}$, picks up an index $a=\pm1$ that specifies whether it is evaluated along forward (``$+$'') or backward (``$-$'') section of the time path.
Accordingly, there are four kinds of two-point functions (``$++$'', ``$--$'', ``$+-$'', and ``$-+$'').
The ``$++$'' and ``$--$'' functions correspond to the usual (anti-)time-ordered propagators,
but the two Wightman functions 
\begin{align}\label{KadanoffBaym}
(\i S_{xy}^{>})_{\alpha\beta} &= \langle \T_C \{ \psi_{x\alpha}^- \overline \psi{}_{y\beta}^+ \} \rangle \ , &
(\i S_{xy}^{<})_{\alpha\beta} &= \langle \T_C \{ \psi_{x\alpha}^+ \overline \psi{}_{y\beta}^- \} \rangle
\end{align}
are the most relevant for our present computation, since they determine the impact of finite particle and antiparticle number densities.
Working with the Fourier-transformed Wightman functions, we use the ansatz
\begin{align}\label{eq:wightman functions}
\i S_p^{<} &= - 2 n(p_0) \, S^{\mathcal A}_p \ , &
\i S_p^{>} &= 2 \left( 1-n(p_0) \right) S_p^{\mathcal A} \ , &
S_{-p}^{\mathcal A} &= \left[ S_p^{\mathcal A} \right]^T \ ,
\end{align}
where $p = (p_0, \bm p)$ is the four-momentum and $S_p^{\mathcal A}$ is the spectral function of $\psi$. 
In thermal equilibrium, the Kubo-Martin-Schwinger relations imply that 
the ansatz \eqref{eq:wightman functions} holds exactly, if
$n(p_0)$ is identified with the Fermi-Dirac distribution $n_F(p_0) = \left[\exp{\beta p_0} + 1\right]^{-1}$.
Allowing the fermion $\psi$ to be out-of-equilibrium, we replace $n_F$ with a generic distribution function, $n_F \to n(p_0, \bm p)$.
In principle, $\psi$ has two independent helicity states, each associated with its own distribution function $n_h(p_0)$, but we neglect this helicity dependence. 
This ansatz is related to the so-called Kadanoff-Baym ansatz, \cf \eg \cite{Garbrecht:2011xw,Drewes:2012qw} for an explicit discussion.
We emphasize that our computation does not depend on the precise shape of the non-equilibrium distribution functions,
provided that they are sufficiently smooth and that the assumptions we make about the spectral function in \cref{sec:spectral function} hold.
As a result, we expect upper bounds akin to those derived in section~\ref{Sec:UpperBound} to hold in out-of-equilibrium scenarios, up to an appropriate modification due to the different phase-space distribution.

We focus on the regime $k \lesssim \mathrm\Upsilon$, where the leading contribution to $\mathrm\Pi$ is generated by the diagram depicted in \cref{fig:wave_diag}.
One way to generate the diagram is to Wick contract the correlator \eqref{eq:stress energy correlator} without introducing any additional vertices.
For a Dirac fermion, this yields
\begin{equation}\begin{aligned}
\left(\frac{4}{c_D}\right)^2 \langle \{ T_{ij} (x), T_{kl} (y) \} \rangle
&= \langle \T_C \left\{
\overline \psi{}_x^- (\i D_{x\, ij} \psi_x^-) \overline \psi{}_y^+ (\i D_{y\, kl} \psi_y^+) 
+ \overline \psi{}_x^+ (\i D_{x\, ij} \psi_x^+) \overline \psi{}_y^- (\i D_{y\, kl} \psi_y^-)
\right\} \rangle \\
&= - \tr[ (\i D_{x\, ij} \i S_{xy}^{>}) (\i D_{y\, kl} \i S_{yx}^{<}) + ( <,> \leftrightarrow >,<) ] \ ,
\end{aligned}\end{equation}
and therefore
\begin{align}
\label{eq:viscosity integral}
\mathrm\Pi(k) &= - \frac{c_D^2}8 \int \frac{\text{d}^3 \bm p}{(2\pi)^3} \hspace{-3pt} \int \frac{\text{d} p_0}{2\pi} \
\mathds L^{ij;kl} p_i p_k \tr[ \gamma_j \i S_{p}^{>} \gamma_l \i S_{p-k}^{<} + \gamma_j \i S_{p}^{<} \gamma_l \i S_{p-k}^{>} ] \ , &
\end{align}
where $k^\mu = (k, \bm k)$.
If $\psi$ is a Majorana fermion, one has to replace $c_D \to c_M = 1$, but the additional contractions result in a further overall prefactor of $2$.
In either case, using the ansatz \eqref{eq:wightman functions}, one obtains the generic expression
\begin{align}
\label{eq:viscosity}
\mathrm\Pi(k) &= c_X \int \frac{\text{d}^3 \bm p}{(2\pi)^3} \hspace{-3pt} \int \frac{\text{d} p_0}{2\pi} \
\mathds L^{ij;kl} p_i p_k \tr[ \gamma_j S_{p^+}^{\mathcal A} \gamma_l S_{p^-}^{\mathcal A} ] \cdot D(p^+_0, p^-_0) \ , &
p^s_\mu &= p_\mu + \frac{s}2 k_\mu \ ,
\end{align}
where \eqref{Ddef} encodes the out-of-equilibrium particle number densities,
\begin{align}
D(p_0, q_0) &= (1-n(p_0)) n(q_0) + n(p_0) (1 - n(q_0)). \nonumber
\end{align}
Using a tree-level or more generic zero-width expression for the spectral function,
the integral in \eqref{eq:viscosity} would vanish exactly for a finite value of $k$ but diverge for $k = 0$.
This divergence is physically related to the fact that a free particle formally has an infinite mean free path,
and is regulated by the finite width of the spectral function, resulting in the $l_\text{av} \propto \nicefrac1{\mathrm\Upsilon}$ scaling of the shear viscosity.
To capture this effect, it is necessary to resum wave-function type corrections to the spectral function.
Following the discussion in appendix \ref{sec:spectral function}, we write the resummed expression as
\begin{align}
\label{eq:spectral function ansatz}
S_p^{\mathcal A} &= \left( \slashed{p} + m \right) \frac{\mathrm\Gamma_p}{\mathrm\Omega_p^2 + \mathrm\Gamma_p^2} \ , &
\mathrm\Omega_p &= p^2 - m^2 \ , & 
\mathrm\Gamma_p &\equiv \mathrm\Gamma(p) = \frac12 \tr\left[ (\slashed{p} + m) \Sigma_p^{\mathcal A} \right] \ ,
\end{align}
where $\Sigma_p^{\mathcal A , \mathcal H}$ denotes the (anti-)hermitian self-energy of $\psi$.
The symmetry relation in \eqref{eq:wightman functions} implies that the width parameter $\mathrm\Gamma_p$ has to be odd, $\mathrm\Gamma_{-p} = - \mathrm\Gamma_p$.
The physical width of the new particle is defined as its on-shell limit,
\begin{align}
\label{eq:generic width}
2 \epsilon \mathrm\Upsilon_p &= \lim_{p_0 \to \epsilon} \mathrm\Gamma_p \ , &
\mathrm\Upsilon_{-p} &= \mathrm\Upsilon_{p} \ , &
\epsilon^2 &= \bm p^2 + m^2 \ .
\end{align}
Inserting the ansatz \eqref{eq:spectral function ansatz} into \cref{eq:viscosity}, one finds%
\footnote{The gamma matrices and metric tensor are defined such that $\{ \gamma_\mu, \gamma_\nu \} = 2 \eta_{\mu\nu}$ with $\eta_{00} = +1$ and $\eta_{ij} = - \delta_{ij}$.}
\begin{align}\label{eq:spectral trace}
\mathds L^{ij;kl} p_i p_k \tr[ \gamma_j S_{p^+}^{\mathcal A} \gamma_l S_{p^-}^{\mathcal A} ]
&= \frac{ 4 p_\perp^2 ( p_0^+ p_0^- - \bm p_+ \bm p_- - m^2 + p_\perp^2) \mathrm\Gamma_{p^+} \mathrm\Gamma_{p^-}}
{ \big[\mathrm\Omega_{p^+}^2 + \mathrm\Gamma_{p^+}^2\big] \big[\mathrm\Omega_{p^-}^2 + \mathrm\Gamma_{p^-}^2 \big] } \ , &
p_\perp^2 &\equiv \mathds L_{ij} p^i p^j \ .
\end{align}
To proceed, we use the residue theorem to evaluate the $p_0$ integral in \eqref{eq:viscosity} in the hydrodynamic regime.
In general, the integral receives contributions from two classes of poles:
(i) poles associated with the product of distribution functions $D = D(p_0^+, p_0^-)$,
and (ii) poles associated with the trace of spectral functions \eqref{eq:spectral trace}.
The width parameter $\mathrm\Gamma$ is also expected to exhibit a branch-cut at the light-cone $p^2 = 0$.
However, for sufficiently small frequencies $k$ and width parameter $\mathrm\Gamma$,
the integral \eqref{eq:viscosity integral} is dominated by the regime in which both propagators are close to being on-shell,
\cf~figures 10 and 11 in \cite{Cheung:2015iqa}%
\footnote{\label{IntegralEnhancement}While figures 10 and 11 in   \cite{Cheung:2015iqa} are useful to illustrate the behaviour of the integral,
there is an error in the computation of the damping that was corrected in \cite{Buldgen:2019dus}.},
where $p^2 > 0$.

As it is the finite width of the spectral function that regulates the integral,
only the residues associated with the spectral functions give rise to the enhanced $\propto \nicefrac1{\mathrm \Upsilon}$ contributions.
Since our primary aim is to extract these enhanced contributions, we neglect the poles of the distribution functions as well as the branch cut of $\mathrm\Gamma_p$,
using an analytic continuation to extend the $p^2 > 0$ expression into the regime with $p^2 < 0$.
We then close the integration contour along the upper half of the complex plane with $\Im p_0 > 0$.
The enhanced contributions are generated by the four poles located at
\begin{align} \label{eq:spectral poles}
p_0 \to \epsilon_{s,t} = - \frac{s k}2 + t \left[ \bm p_s^2 + m^2 + \left\{\i \mathrm\Gamma_{p^s} \right\}_{p_0 \to \epsilon_{s,t}} \right]^{\frac12} \ ,
\end{align}
where $s,t = \pm 1$.
The corresponding residues are
\begin{align} \label{eq:spectral residues}
\mathop{\text{Res}}_{p_0 \to \epsilon_{s,t}} \left( \frac{\mathrm\Gamma_{p^+}\mathrm\Gamma_{p^-}}
{ \big[\mathrm\Omega_{p^{+}}^2 + \mathrm\Gamma_{p^+}^2 \big] \big[\mathrm\Omega_{p^{-}}^2 + \mathrm\Gamma_{p^-}^2 \big] } \right)
&= \lim_{p_0 \to \epsilon_{s,t}} \frac{\mathrm\Gamma_{(p^s - s k)}}
{2\i \big(2 \epsilon_{s,t} + s k - \i \mathrm\Gamma_{p^s}^\prime \big) \big[ (2 s k^\mu p_\mu^s - \i \mathrm\Gamma_{p^s})^2 + \mathrm\Gamma_{(p^s - s k)}^2 \big] } \ ,
\end{align}
where $\mathrm\Gamma_p^\prime \equiv \partial_{p_0} \mathrm\Gamma_p$ is an even function of $p_0$.
To evaluate \cref{eq:spectral residues}, we separately expand both the numerator and denominator to leading order in $k$ and $\mathrm\Gamma_p$.
Explicitly, this gives
\begin{align}\label{eq:soft approximations}
\mathrm\Gamma_{p^s} \approx \mathrm\Gamma_{(p^s - sk)} &\approx \mathrm\Gamma_p \ , &
\epsilon_{s,t} &\approx t \epsilon + \i \mathrm\Upsilon_{p} - s k \frac{(\epsilon - t p_\sparallel)}{2\epsilon} \ , &
k^\mu p^s_\mu &\approx k (t \epsilon - p_\sparallel) \ ,
\end{align}
where $p_\sparallel = \flatfrac{\bm k \bm p}k$, and subsequently
\begin{align} \label{eq:spectral residue approx}
\mathop{\text{Res}}_{p_0 \to \epsilon_{s,t}} \left( \frac{\mathrm\Gamma_{p^+}\mathrm\Gamma_{p^-}}
{ \big[\mathrm\Omega_{p^{+}}^2 + \mathrm\Gamma_{p^+}^2 \big] \big[\mathrm\Omega_{p^{-}}^2 + \mathrm\Gamma_{p^-}^2 \big] } \right)
&\approx \frac1{8\i} \frac{\mathrm\Upsilon_p} {[ s k (t \epsilon - p_\sparallel) - \i t \epsilon \mathrm\Upsilon_p]^2 + \epsilon^2 \mathrm\Upsilon_p^2 } \ .
\end{align}
Hence, summing over the residues and setting $k\to0$ and $\epsilon_{s,t} \to t \epsilon$ in the remainder of the integrand in \eqref{eq:viscosity} gives \eqref{GeneralPi}.

The result \eqref{eq:fermionic prodrate} is a valid approximation of the wave-function type diagram in \cref{fig:wave_diag}.
This should hold as long as the integral is dominated by the dressed single quasiparticle poles,
which can be expected for momenta $k \ll \mathrm\Upsilon_p$, provided that the phase space distributions do not feature pronounced peaks.
We also expect it to work reasonably well for larger momenta $\mathrm\Upsilon_p < k \ll T$.
The neglected contributions are generated by 
additional poles of the spectral function (\eg related to collective excitations \cite{Klimov:1982bv,Weldon:1982bn}),
the poles of the distribution function $D(p_0^+, p_0^-)$, and the branch cut of the width parameter $\mathrm\Gamma_p$.
The residues of the distribution function generically scale as $\mathrm\Upsilon_p^2$, and are therefore in any case subleading compared to the residues \eqref{eq:spectral residues}.
We expect that a similar suppression is present for the branch cut contribution at sufficiently small momenta, but this may no longer be the case for momenta $k \gg \mathrm\Upsilon_p$.
Furthermore, for momenta $k\sim \pi T$, we can no longer use the approximations \eqref{eq:soft approximations} or set $k\to0$ in the non-singular part of the integrand \eqref{eq:viscosity}.
It is then also necessary to account for additional diagrams such as vertex-type corrections.
Finally, we note that the equations used here account for the production of single gravitons, and do no cover the production of coherent multi-graviton states.
Emission of multiple hard gravitons with momenta  $k \sim \pi T$ was studied in \cite{Ghiglieri:2022rfp}, while emission of multiple soft gravitons could \eg be included by studying the evolution of higher $n$-point functions.

\section{Finite temperature spectral function}
\label{sec:spectral function}

We consider the spectral function of a massive Dirac or Majorana fermion $\psi$ at finite temperature,
\begin{align}
S_p^{\mathcal A} &\equiv
\frac12 \int \frac{\text{d}^4 p}{(2\pi)^4} e^{\i p(x-y)} \langle \psi_x \otimes \overline \psi_y + \overline \psi_y \otimes \psi_x \rangle \ ,
\end{align}
where $\otimes$ denotes the tensor product in Dirac space.
In thermal equilibrium, an exact, if formal, expression is
\begin{align}\label{SpectralFull}
S_p^{\mathcal A} &= \frac{\i}2 \left( S_p^{r} - S_p^{a} \right) =
\frac{\i}{2} \left( \frac{1}{\slashed p - m - \Sigma^{\mathcal H}_p + \i \Sigma^{\mathcal A}_p}
- \frac{1}{\slashed p - m - \Sigma^{\mathcal H}_p - \i \Sigma^{\mathcal A}_p} \right) \ ,
\end{align}
where $S_p^{r}$ and $S_p^{a}$ are the retarded and advanced propagators of $\psi$ and $\Sigma_p^{\mathcal H}$ and $\Sigma_p^{\mathcal A}$ are its (anti-)hermitian self-energies.
Working in the plasma rest frame, they can be cast as%
\begin{align}
\label{eq:self energies}
\Sigma_{p}^{\mathcal X} &=
\frac{ \left(\sigma_1^{\mathcal X} \, \slashed{p} + \tilde \sigma_1^{\mathcal X} \, \slashed{\tilde p}\right) \mathds 1
+ \left(\sigma_5^{\mathcal X} \, \slashed{p} + \tilde \sigma_5^{\mathcal X} \, \slashed{\tilde p}\right) \gamma_5}{p^2}
+ \delta_1^{\mathcal X} \, \mathds 1 + \delta_5^{\mathcal X} \gamma_5 \ , &
\tilde p^\mu &= (\abs{\bm p}, \nicefrac{p_0 \bm p}{\abs{\bm p}} ) \ ,
\end{align}
where $\mathcal X \in \left\{\mathcal A, \mathcal H \right\}$.
The quantity $\tilde p^\mu$ is defined such that $\tilde p^2 = - p^2$ and $\tilde p p = 0$.
This ensures that $p^\mu$ and $\tilde p^\mu$ are linearly independent but span the space of four-vectors with spatial part parallel to $\bm p$.
The scalar coefficients are
\begin{equation}\label{eq:loop coefficients}
\begin{aligned}
\sigma_i^{\mathcal X} &= \frac14 \tr \left[ \P_i \slashed p \, \Sigma^{\mathcal X}_p \right] \ , &
\tilde \sigma_i^{\mathcal X} &= - \frac14 \tr \left[ \P_i \slashed{\tilde p} \, \Sigma^{\mathcal X}_p \right] \ , &
\delta_i^{\mathcal X} &= \frac14 \tr \left[ \P_i \Sigma^{\mathcal X}_p \right] \ , &
\P_5 &= \gamma_5 \ , &
\P_1 &= \mathds 1 \ .
\end{aligned}\end{equation}
Using the decomposition \eqref{eq:self energies}, we can recast the retarded and advanced propagators as
\begin{align}
\label{eq: advanced retarded propagators}
S_p^{a,r}
&= \left( \slashed{p} + m - \overline \Sigma^{\mathcal H}_p \mp \i \overline \Sigma^{\mathcal A}_p \right) \frac1{\mathrm\Omega_p \mp \i \mathrm\Gamma_p} \ , &
\overline \Sigma^{\mathcal X}_p &= \Sigma^{\mathcal X}_p - 2 \delta_1^{\mathcal X} \mathds 1 \ ,
\end{align}
where the functions $\mathrm\Omega_p$ and $\mathrm\Gamma_p$ are defined as
\begin{subequations}\begin{align}
p^2 \mathrm \Omega_p &\equiv 
\begin{multlined}[t]
(p^2- \sigma_1^{\mathcal H})^2 - \sigma_1^{\mathcal A \, 2}
- \tilde \sigma_1^{\mathcal H \, 2} + \tilde \sigma_1^{\mathcal A \, 2}
- \sigma_5^{\mathcal H \, 2} + \sigma_5^{\mathcal A \, 2}
+ \tilde \sigma_5^{\mathcal H \, 2} - \tilde \sigma_5^{\mathcal A \, 2} \\
- p^2 [ (m+\delta_1^{\mathcal H})^2 - \delta_1^{\mathcal A \ , 2} - \delta_5^{\mathcal H \, 2} + \delta_5^{\mathcal A \, 2} ]
\ , 
\end{multlined}
\\
p^2 \mathrm \Gamma_p &\equiv
2 \left[ (p^2 - \sigma_1^{\mathcal H}) \sigma_1^{\mathcal A}
+ \tilde \sigma_1^{\mathcal H} \tilde \sigma_1^{\mathcal A}
+ \sigma_5^{\mathcal H} \sigma_5^{\mathcal A}
- \tilde \sigma_5^{\mathcal H} \tilde \sigma_5^{\mathcal A} \right]
+ 2 p^2 [ \delta_1^{\mathcal A} (m + \delta_1^{\mathcal H}) - \delta_5^{\mathcal H} \delta_5^{\mathcal A}] \ .
\end{align}\end{subequations}
The spectral function becomes
\begin{align}
\label{eq:HNL resummed spectral}
S_{p}^{\mathcal A}
&= \left( \slashed{p} + m - \overline \Sigma^{\mathcal H}_p \right) \Delta_p - \overline \Sigma^{\mathcal A}_p \Pi_p \ ,
\end{align}
where
\begin{align}
\Delta_p &= \frac{\mathrm\Gamma_p}{\mathrm\Omega_p^2 + \mathrm\Gamma_p^2} \ , &
\Pi_p &= \frac{\mathrm\Omega_p}{\mathrm\Omega_p^2 + \mathrm\Gamma_p^2} \ .
\end{align}
The condition $\mathrm \Omega_p \overset{!}{=} 0$ determines the location of the mass shell,
while $4 \mathrm\Gamma_p = \tr\left[ (\slashed p + m) \Sigma_p^{\mathcal A} \right]$ is proportional to the finite temperature width of the particle.
In the zero-width approximation, where $\mathrm\Gamma_p \to 0$, one finds
\begin{align}
\Delta_p &\to \pi \delta(\Omega_p) \ , &
\Pi_p &\to \P \left( \frac1{\Omega_p} \right) \ ,
\end{align}
where $\P \left( \cdot \right)$ denotes the principal value.
In this limit, the term $\propto \Delta_p$ encodes the propagation of single-particle excitations,
encompassing both pseudo-particle and hole modes, while the term $\propto \Pi_p$ encodes the impact of scatterings mediated by these modes.
In our computation, we can expand $S_p^{\mathcal A}$, $\mathrm \Omega$, and $\mathrm \Gamma$ to leading order in the coefficients \eqref{eq:loop coefficients},
\begin{align}
S_{p}^{\mathcal A} &= (\slashed{p} + m) \Delta_p + \order{\Sigma^2} \ , &
\mathrm \Omega &= p^2 - m^2 \ , & 
\mathrm \Gamma &= 2 \left( \sigma_1^{\mathcal A} + m \, \delta_1^{\mathcal A} \right) + \order{\Sigma^2} \ .
\end{align}
It is worth pointing out that this approximation, neglecting loop corrections to $\mathrm\Omega$ as well as the Dirac structure of the spectral function, is not always valid.
In particular, for soft momenta $\abs{\bm p} \ll T$, one has to at least keep track of the hard-thermal loop contributions to $\mathrm\Omega$
in order to regulate soft IR divergences and to reproduce the correct spectrum of the fermion propagator, including both massive pseudoparticle and hole modes.
In our case, the integral \eqref{eq:viscosity} is dominated by hard momenta $\abs{\bm p} \sim \pi T$ that are close to the lightcone, such that $p^2 \ll T^2$.
In this regime, the residue associated with the hole-modes is exponentially suppressed, and they can be neglected as a result,
but the pseudoparticle modes still obtain a thermal mass $\propto g T$ that regulates IR divergences and can therefore result in certain logarithmically enhanced contributions.
This can be accounted for by promoting $m$ to a temperature dependent thermal mass, $m\to m(T)$, but it does not impact the validity or our computation.
In addition, since there are no more IR divergences to regulate, keeping track of the thermal mass barely modifies our final result \eqref{eq:fermionic prodrate}.

\section{Supplementary formulae for backgrounds from\\ non-renormalizable interactions}
\label{sec:gwb appendix}

To estimate the final \GWB \eqref{eq:final gw background} from hidden particles with a width dominated by non-renormalizable interactions,
we decompose the high-temperature production rate \eqref{eq:relat rate} into a hydrodynamic ($T > T_c$) and Boltzmann ($T < T_c$) contribution,
\begin{align}
\mathrm\Pi(k) &\overset{m \ll T}{\simeq} g_X \frac{16 \pi^2}{225} T^5
\begin{cases}
\nicefrac1{2 \mathrm\Upsilon_\text{av}} & T > T_c \\
\nicefrac{5 \mathrm\Upsilon_\text{av}}{k^2} & T < T_c
\end{cases}
\ .
\end{align}
We extract the high-temperature background from the integral in \eqref{eq:final gw background} by replacing $T_\text{min}$ with $m$.
Using expression \eqref{eq:width scaling} for the width $\mathrm\Upsilon_\text{av}$ and assuming $d > 4$, one obtains the hydrodynamic contribution
\begin{align}
h^2 \Omega_\text{gw}^{T > m}(f)
&\overset{\text{Hydro}}{\simeq} 
g_d^- \left( \frac{f}{\unit{Hz}} \right)^2 
\left[ \min\left( \frac{f \, m}{f_\text{hyd} \, m_\text{Pl}}, \frac{T_c}{m_\text{Pl}} \right)
- \min\left( \frac{f \, T_\text{max}}{f_\text{max} \, m_\text{Pl}} , \frac{T_c}{m_\text{Pl}} \right) \right]
\end{align}
and the Boltzmann contribution
\begin{align}
h^2 \Omega_\text{gw}^{T > m}(f)
&\overset{\text{Boltz.}}{\simeq} 
g_d^+ \left( \frac{f}{\unit{Hz}} \right)^2 
\left[ \min \left( \frac{f_\text{max} \, T_\text{max}}{f \, m_\text{Pl}}, \frac{T_c}{m_\text{Pl}} \right)
- \min \left( \frac{f_\text{hyd} \, m}{f \, m_\text{Pl}}, \frac{T_c}{m_\text{Pl}} \right) \right] \ ,
\end{align}
where
\begin{align}\label{eq:fmax def}
g_d^\pm &= \frac{5.3 \, g_X \cdot 10^{-29}}{2(d-4) \pm 1} \ , &
T_\text{max} &= \max\left(m, \min \left( T_\text{entry}, T_\star \right) \right) \ , &
f_\text{max} &= f_c \left( \frac{T_\text{max}}{\Lambda} \right)^{2(d-4)} \ .
\end{align}
Adding these two contributions, one obtains the full background
\begin{align}
h^2 \Omega_\text{gw}^{T > m}(f) &\simeq 
g_d^+ \left( \frac{f}{\unit{Hz}} \right)^2
\begin{cases}
\frac{g_d^-}{g_d^+} \left( \frac{f \, m}{f_\text{hyd} m_\text{Pl}} - \frac{f \, T_\text{max}}{f_\text{max} \, m_\text{Pl}} \right) & f_\text{hyd} > f \\
\left( \frac{T_c}{m_\text{Pl}} - \frac{f_\text{hyd} \, m}{f \, m_\text{Pl}} \right) + \frac{g_d^-}{g_d^+} \left( \frac{T_c}{m_\text{Pl}} - \frac{f \, T_\text{max}}{f_\text{max} \, m_\text{Pl}} \right)
& f_\text{hyd} < f < f_\text{max} \\
\left( \frac{f_\text{max} \, T_\text{max}}{f \, m_\text{Pl}} - \frac{f_\text{hyd} \, m}{f \, m_\text{Pl}} \right) & f_\text{max} < f
\end{cases} \ .
\end{align}
Keeping only the leading terms for each case, one has
\begin{align}
h^2 \Omega_\text{gw}^{T > m}(f) &\simeq 
g_d^+ \left( \frac{f}{\unit{Hz}} \right)^2
\begin{cases}
\frac{g_d^-}{g_d^+} \frac{f \, m}{f_\text{hyd} m_\text{Pl}} & f_\text{hyd} > f \\
\left( 1 + \frac{g_d^-}{g_d^+} \right) \frac{T_c}{m_\text{Pl}} 
& f_\text{hyd} < f < f_\text{max} \\
\frac{f_\text{max} \, T_\text{max}}{f \, m_\text{Pl}}  & f_\text{max} < f
\end{cases} \ .
\end{align}
In the following, we focus on the case $m < T_\star < \Lambda$.
From the definitions \eqref{eq:fstar def,eq:cutoff frequency,eq:fhyd def,eq:fbol def,eq:fcross def},
it then follows that $0 < f_\star - f_\text{nr} < f_\text{bol} - f_\text{hyd}$.
From \eqref{eq:fmax def}, we also find
\begin{align}
f_\text{max} &= 
\begin{cases}
f_\text{hyd} & f < f_\text{nr} \\
f_\text{int} & f_\text{nr} < f < f_\star \\
f_\text{bol} & f_\star < f
\end{cases}\ , &
f_\text{int} &\equiv f_x \left( \frac{f}{f_x} \right)^{2(d-4)} \ ,
\end{align}
where $f_x$ is defined in \eqref{eq:fcross def}, and
\begin{align}
f_\text{hyd} &< f_\text{max} < f_\text{bol} \ , &
f_\text{hyd} &= f_x \left( \frac{f_\text{nr}}{f_x} \right)^{2(d-4)} \ , &
f_\text{bol} &= f_x \left( \frac{f_\star}{f_x} \right)^{2(d-4)} \ .
\end{align}
For $f < f_\text{nr}$, the temperature at horizon entry is below the mass of the hidden particle, so that there is no high-temperature ($T>m$) contribution to the background.
Focusing on frequencies $f > f_\text{nr}$, we now distinguish four cases:
\begin{itemize}
\item
Case 1:
$f_\text{nr} < f_\text{hyd} < f_\star < f_\text{bol}$,
which gives
\begin{align}
h^2 \Omega_\text{gw}^{T > m}(f) &\simeq 
g_d^+ \left( \frac{f}{\unit{Hz}} \right)^2
\begin{cases}
\frac{g_d^-}{g_d^+} \left( \frac{f \, m}{f_\text{hyd} m_\text{Pl}} - \frac{f \, T_\text{entry}}{f_\text{int} \, m_\text{Pl}} \right) & f_\text{nr} < f < f_\text{hyd} \\
\left( \frac{T_c}{m_\text{Pl}} - \frac{f_\text{hyd} \, m}{f \, m_\text{Pl}} \right) + \frac{g_d^-}{g_d^+} \left( \frac{T_c}{m_\text{Pl}} - \frac{f \, T_\text{entry}}{f_\text{int} \, m_\text{Pl}} \right)
& f_\text{hyd} < f < f_\star \\
\left( \frac{T_c}{m_\text{Pl}} - \frac{f_\text{hyd} \, m}{f \, m_\text{Pl}} \right) + \frac{g_d^-}{g_d^+} \left( \frac{T_c}{m_\text{Pl}} - \frac{f \, T_\star}{f_\text{bol} \, m_\text{Pl}} \right)
& f_\star < f < f_\text{bol} \\
\left( \frac{f_\text{bol} \, T_\star}{f \, m_\text{Pl}} - \frac{f_\text{hyd} \, m}{f \, m_\text{Pl}} \right) & f_\text{bol} < f
\end{cases} \ .
\end{align}
Keeping only the leading terms, one has
\begin{align}\label{eq:nonrel gw case 1 leading}
h^2 \Omega_\text{gw}^{T > m}(f) &\simeq 
g_d^+ \left( \frac{f}{\unit{Hz}} \right)^2
\begin{cases}
\frac{g_d^-}{g_d^+} \frac{f \, m}{f_\text{hyd} m_\text{Pl}} & f_\text{nr} < f < f_\text{hyd} \\
\left( 1 + \frac{g_d^-}{g_d^+} \right) \frac{T_c}{m_\text{Pl}} 
& f_\text{hyd} < f < f_\text{bol} \\
\frac{f_\text{bol} \, T_\star}{f \, m_\text{Pl}}& f_\text{bol} < f
\end{cases} \ .
\end{align}
\item
Case 2:
$f_\text{nr} < f_\star < f_\text{hyd} < f_\text{bol}$,
which gives
\begin{align}
h^2 \Omega_\text{gw}^{T > m}(f) &\simeq 
g_d^+ \left( \frac{f}{\unit{Hz}} \right)^2
\begin{cases}
\frac{g_d^-}{g_d^+} \left( \frac{f \, m}{f_\text{hyd} m_\text{Pl}} - \frac{f \, T_\text{entry}}{f_\text{int} \, m_\text{Pl}} \right) & f_\text{nr} < f < f_\star \\
\frac{g_d^-}{g_d^+} \left( \frac{f \, m}{f_\text{hyd} m_\text{Pl}} - \frac{f \, T_\star}{f_\text{bol} \, m_\text{Pl}} \right) & f_\star < f < f_\text{hyd} \\
\left( \frac{T_c}{m_\text{Pl}} - \frac{f_\text{hyd} \, m}{f \, m_\text{Pl}} \right) + \frac{g_d^-}{g_d^+} \left( \frac{T_c}{m_\text{Pl}} - \frac{f \, T_\star}{f_\text{bol} \, m_\text{Pl}} \right)
& f_\text{hyd} < f < f_\text{bol} \\
\left( \frac{f_\text{bol} \, T_\star}{f \, m_\text{Pl}} - \frac{f_\text{hyd} \, m}{f \, m_\text{Pl}} \right) & f_\text{bol} < f
\end{cases} \ .
\end{align}
Keeping only the leading terms, one has
\begin{align}
h^2 \Omega_\text{gw}^{T > m}(f) &\simeq 
g_d^+ \left( \frac{f}{\unit{Hz}} \right)^2
\begin{cases}
\frac{g_d^-}{g_d^+} \frac{f \, m}{f_\text{hyd} m_\text{Pl}} & f_\text{nr} < f < f_\text{hyd} \\
\left( 1 + \frac{g_d^-}{g_d^+} \right) \frac{T_c}{m_\text{Pl}}
& f_\text{hyd} < f < f_\text{bol} \\
\frac{f_\text{bol} \, T_\star}{f \, m_\text{Pl}} & f_\text{bol} < f
\end{cases} \ .
\end{align}
Notice that this expression is the same as \cref{eq:nonrel gw case 1 leading}.
\item
Case 3:
$f_\text{hyd} < f_\text{nr} < f_\star < f_\text{bol}$,
which gives
\begin{align}
h^2 \Omega_\text{gw}^{T > m}(f) &\simeq 
g_d^+ \left( \frac{f}{\unit{Hz}} \right)^2
\begin{cases}
\left( \frac{f_\text{int} \, T_\text{entry}}{f \, m_\text{Pl}} - \frac{f_\text{hyd} \, m}{f \, m_\text{Pl}} \right) & f_\text{nr} < f < f_x \\
\left( \frac{T_c}{m_\text{Pl}} - \frac{f_\text{hyd} \, m}{f \, m_\text{Pl}} \right) + \frac{g_d^-}{g_d^+} \left( \frac{T_c}{m_\text{Pl}} - \frac{f \, T_\text{entry}}{f_\text{int} \, m_\text{Pl}} \right)
& f_x < f < f_\star \\
\left( \frac{T_c}{m_\text{Pl}} - \frac{f_\text{hyd} \, m}{f \, m_\text{Pl}} \right) + \frac{g_d^-}{g_d^+} \left( \frac{T_c}{m_\text{Pl}} - \frac{f \, T_\star}{f_\text{bol} \, m_\text{Pl}} \right)
& f_\star < f < f_\text{bol} \\
\left( \frac{f_\text{bol} \, T_\star}{f \, m_\text{Pl}} - \frac{f_\text{hyd} \, m}{f \, m_\text{Pl}} \right) & f_\text{bol} < f
\end{cases} \ .
\end{align}
Keeping only the leading terms, one has
\begin{align}
h^2 \Omega_\text{gw}^{T > m}(f) &\simeq 
g_d^+ \left( \frac{f}{\unit{Hz}} \right)^2
\begin{cases}
\frac{f_\text{int} \, T_\text{entry}}{f \, m_\text{Pl}} & f_\text{nr} < f < f_x \\
\left( 1 + \frac{g_d^-}{g_d^+} \right) \frac{T_c}{m_\text{Pl}}
& f_x < f < f_\text{bol} \\
\frac{f_\text{bol} \, T_\star}{f \, m_\text{Pl}} & f_\text{bol} < f
\end{cases} \ .
\end{align}
\item
Case 4:
$f_\text{hyd} < f_\text{nr}, f_\text{bol} < f_\star$,
which gives
\begin{align}
h^2 \Omega_\text{gw}^{T > m}(f) &\simeq 
g_d^+ \left( \frac{f}{\unit{Hz}} \right)^2
\begin{cases}
\left( \frac{f_\text{int} \, T_\text{entry}}{f \, m_\text{Pl}} - \frac{f_\text{hyd} \, m}{f \, m_\text{Pl}} \right) & f_\text{nr} < f < f_\star \\
\left( \frac{f_\text{bol} \, T_\star}{f \, m_\text{Pl}} - \frac{f_\text{hyd} \, m}{f \, m_\text{Pl}} \right) & f_\star < f
\end{cases} \ .
\end{align}
Keeping only the leading terms, one has
\begin{align}
h^2 \Omega_\text{gw}^{T > m}(f) &\simeq 
g_d^+ \left( \frac{f}{\unit{Hz}} \right)^2
\begin{cases}
\frac{f_\text{int} \, T_\text{entry}}{f \, m_\text{Pl}}  & f_\text{nr} < f < f_\star \\
\frac{f_\text{bol} \, T_\star}{f \, m_\text{Pl}} & f_\star < f
\end{cases} \ .
\end{align}
\end{itemize}
Considering the limit of a light hidden particle, where $f_\text{hyd}, f_\text{nr} \to 0$, only two cases survive:
\begin{itemize}
\item
Case 3, which gives
\begin{align}
h^2 \Omega_\text{gw}^{T > m}(f) &\simeq 
g_d^+ \left( \frac{f}{\unit{Hz}} \right)^2
\begin{cases}
\frac{f_\text{int} \, T_\text{entry}}{f \, m_\text{Pl}} & f < f_x \\
\frac{T_c}{m_\text{Pl}} + \frac{g_d^-}{g_d^+} \left( \frac{T_c}{m_\text{Pl}} - \frac{f \, T_\text{entry}}{f_\text{int} \, m_\text{Pl}} \right)
& f_x < f < f_\star \\
\frac{T_c}{m_\text{Pl}} + \frac{g_d^-}{g_d^+} \left( \frac{T_c}{m_\text{Pl}} - \frac{f \, T_\star}{f_\text{bol} \, m_\text{Pl}} \right)
& f_\star < f < f_\text{bol} \\
\frac{f_\text{bol} \, T_\star}{f \, m_\text{Pl}} & f_\text{bol} < f
\end{cases} \ .
\end{align}
\item
Case 4, which gives
\begin{align}
h^2 \Omega_\text{gw}^{T > m}(f) &\simeq 
g_d^+ \left( \frac{f}{\unit{Hz}} \right)^2
\begin{cases}
\frac{f_\text{int} \, T_\text{entry}}{f \, m_\text{Pl}} & f < f_\star \\
\frac{f_\text{bol} \, T_\star}{f \, m_\text{Pl}} & f_\star < f
\end{cases} \ .
\end{align}
\end{itemize}

\section{More general cosmic histories}
\label{sec:non standard temperature evolution appendix}

Integrating the graviton equation of motion \eqref{eq:gravitational wave eom}, and using the numerical constants in \eqref{eq:photon energy density numbers}, one has
\begin{align}\label{eq:general gw spectrum}
h^2 \Omega_\text{gw} (f)
&\approx 1.31 \cdot 10^{-37} \times \left( \frac{f}{\unit{Hz}} \right)^3 \int\displaylimits_{t_i}^{t_f} \hspace{-3pt} \text{d} t \ \frac{a}{a_0 T_0} \frac{\mathrm\Pi(\flatfrac{2\pi f a_0}a)}{m_\text{Pl}^2}
\ ,
\end{align}
where $t_i$ is the initial time at the onset and $t_f$ the final time at the end of \GW production.
In order to evaluate the integral, one has to supply expressions for $a$ and $\mathrm \Pi$ as functions of time.
In \cref{sec:present spectrum}, we assumed that the \SM is in kinetic equilibrium, that it dominates the overall energy budget of the universe, and that there is no entropy exchange with other sectors.
If the \SM does not dominate the energy budget but the other two assumptions do remain valid, one obtains
\begin{align}
h^2 \Omega_\text{gw}(f)
&\approx 
2.02 \cdot 10^{-38} \times \left( \frac{f}{\unit{Hz}} \right)^3
\times \hspace{-3pt} \int\displaylimits_{T_\text{min}}^{T_\text{max}} \hspace{-3pt} \frac{\text{d} T^\prime}{m_\text{Pl}}
\frac{\mathrm\Pi \left( \flatfrac{2\pi f a_0}{a^\prime} \right)}{8T^{\prime 4}} \, 
\left( \frac{\rho_\text{SM}}{\rho_\text{tot}} \right)^{\frac12}
\ ,
\end{align}
generalizing \cref{eq:gravitational wave spectrum}.
This expression is strictly smaller than \cref{eq:gravitational wave spectrum}, so that it does not affect the upper bounds derived in \cref{Sec:UpperBound}.
Of course, this does not imply that the \GWB is always smaller than in cases where the \SM dominates the energy budget,
since the impact of new particles may also enhance the production rate $\mathrm\Pi \left( \flatfrac{2\pi f a_0}{a^\prime} \right)$, while still respecting the upper bounds in \cref{Sec:UpperBound}.

If significant amounts of entropy are transferred between the \SM and other sectors, it becomes necessary to solve the Friedmann equations in order to directly obtain the time-evolution of the \SM temperature $T$ and the Hubble rate $H$.
If $T$ and $H$ are known monotonous functions of the scale-factor, one can rewrite \cref{eq:general gw spectrum} as
\begin{align}\label{eq:finite entropy transfer gw spectrum}
h^2 \Omega_\text{gw} (f)
&\approx 2.02 \cdot 10^{-38} \times \left( \frac{f}{\unit{Hz}} \right)^3 \int\displaylimits_{T_\text{min}}^{T_\text{max}} \hspace{-3pt} \frac{\text{d} T^\prime}{m_\text{Pl}} \ 
\frac{\mathrm\Pi(\flatfrac{2\pi f a_0}a)}{8 T^{\prime 4}}
\left| \frac{\text{d}\ln a}{\text{d} \ln T^\prime} \right|
\left(\frac{T^\prime}{\overline T^\prime} \right)
\left( \frac{\rho_\text{SM}}{\rho_\text{tot}} \right)^{\frac12}
\ ,
\end{align}
where
\begin{align}
\overline T \equiv \frac{a_0 T_0}{a} \left[ \frac{g_s^\text{SM}(T_0)}{g_s^\text{SM}(\overline T)} \right]^{\nicefrac13}
\end{align}
is what the  \SM temperature would be in the absence of entropy transfers.
Using the \SM production rate \eqref{eq:sm gw production rate}, one obtains the final background from \SM physics
\begin{align}\label{eq:finite entropy transfer gw spectrum SM}
h^2 \Omega_\text{gw}^\text{SM} (f)
&\approx 1.03 \cdot 10^{-36} \times \left( \frac{f}{\unit{Hz}} \right)^3 \int\displaylimits_{T_\text{min}}^{T_\text{max}} \hspace{-3pt} \frac{\text{d} T^\prime}{m_\text{Pl}} \
\left| \frac{\text{d}\ln a}{\text{d} \ln T^\prime} \right|
\left(\frac{T^\prime}{\overline T^\prime} \right)
\left( \frac{\rho_\text{SM}}{\rho_\text{tot}} \right)^{\frac12}
\ ,
\end{align}
which extends \eqref{eq:final gw background} to more general cosmic histories.
Note that the product
$\left|\frac{\text{d}\ln a}{\text{d} \ln T^\prime}\right|\frac{T^\prime}{\overline T^\prime} $
in \eqref{eq:finite entropy transfer gw spectrum} also makes it possible to describe \GWB production without entropy transfer between sectors
during periods with an expansion history that deviated from radiation domination, which can e.g.~be realised during reheating.

\printbibliography

\end{document}